\documentclass[fleqn,10pt]{wlscirep}
\usepackage[utf8]{inputenc}
\usepackage[T1]{fontenc}
\usepackage{bm}
\usepackage{prettyref}
\usepackage{lineno}
\usepackage{comment}
\usepackage[normalem]{ulem}

\let\mathcal\undefined 
\DeclareMathAlphabet{\mathcal}{OMS}{cmsy}{m}{n}
\newrefformat{eq}{(\ref{#1})} 

\newcommand{\ii}{\mathrm{i}}
\newcommand{\ee}{\mathrm{e}}
\newcommand{\JJ}{\mathit{J}}

\title{Frequency Selective Wave Beaming in Nonreciprocal Acoustic Phased Arrays}

\author[1,+]{Revant Adlakha}
\author[2,+]{Mohammadreza Moghaddaszadeh}
\author[1,+]{Mohammad A. Attarzadeh}
\author[2]{Amjad Aref}
\author[1,*]{Mostafa Nouh}
\affil[1]{Department of Mechanical and Aerospace Engineering, University at Buffalo, Buffalo, NY 14260, USA}
\affil[2]{Department of Civil, Structural and Environmental Engineering, University at Buffalo, Buffalo, NY 14260, USA}
\affil[+]{these authors contributed equally to this work}
\affil[*]{Corresponding author: \url{mnouh@buffalo.edu}}

\begin{abstract}
Acoustic phased arrays are capable of steering and focusing a beam of sound via selective coordination of the spatial distribution of phase angles between multiple sound emitters. Constrained by the principle of reciprocity, conventional phased arrays exhibit identical transmission and reception patterns which limit the scope of their operation. This work presents a controllable space-time acoustic phased array which breaks time-reversal
symmetry, and enables phononic transition in both momentum and energy spaces. By leveraging a dynamic phase modulation, the proposed linear phased array is no longer bound by the acoustic reciprocity, and supports asymmetric transmission and reception patterns that can be tuned independently at multiple channels. A foundational framework is developed to characterize and interpret the emergent nonreciprocal phenomena and is later validated against benchmark numerical experiments. The new phased array selectively alters the directional and frequency content of the incident signal and the frequency conversion between the different wave fields is analyzed as a function of the imposed modulation. The space-time acoustic phased array enables unprecedented control over sound waves in a variety of applications ranging from ultrasonic imaging to non-destructive testing and underwater SONAR telecommunication.
\end{abstract}

\begin{document}
\flushbottom
\maketitle
\thispagestyle{empty}

\section*{Introduction}

In their most general form, phased arrays can be thought of as a coalescence of multiple wave transmitting/receiving components---also known as transceivers---which share a common excitation/collection port. The hallmark feature of phased arrays, setting them apart from antennas, is an additionally imparted phase angle on each of its individual transceivers. The ability to manipulate an incident wavefront, made possible by such phase variations, breeds new opportunities in beam focusing and guidance as well as the capability to efficiently receive a signal from an arbitrary direction \cite{beamsteering, beamsteering1}. Phased arrays were first proposed for military use to quickly scan a sky range via electromagnetic waves in search of flying objects, replacing bulky mechanically-rotating antennas which served the same function \cite{jones1988technical}. Owing to their ability to steer beams, they quickly infiltrated a wide range of civil applications in optics, ultrasonics, and acoustics. Recent examples include LIDAR \cite{poulton_lidar}, RADAR \cite{radar}, SONAR \cite{jorgensen1993doppler}, medical ultrasound imaging \cite{jensen2007medical}, geology and seismology \cite{rynne1998phased,birtill1965application} as well as Non-Destructive Testing (NDT) \cite{mcnab1987ultrasonic}. Ultrasonic phased arrays have been employed for obstacle detection, depth measurement, as well as NDT mechanisms to identify defects in composite-stiffened structures \cite{zhou_2016}. Another emerging application is acoustic levitation, where phased arrays were utilized to create standing waves and trap a particle at pressure nodes \cite{marzo_virtualvortex,hoshi_2014}. Using a similar configuration, acoustically controlled holograms have been most recently reported \cite{marzo_tweezers}. 
Nowadays, phased arrays are being used in the development of the SpaceX Starlink constellation to enhance global internet connectivity by exploiting its beam forming properties \cite{starlink1, starlink2}. 
They have also been explored to enhance wireless capabilities of in-home WiFi and cellular networks \cite{bashri, qian}. 
Additionally, phased arrays have found applications in weather forecasting \cite{weather}, astronomy and interstellar communication \cite{arnon:s}, among others.

Depending on their geometric configuration, phased arrays are categorized as planar or in-line arrangements.
By virtue of their sub-wavelength nature, a planar phased array is sufficient to effectively shape wave beams in a 3-Dimensional (3D) space; an in-line arrangement is capable of the same in a 2D space.
As such, phased arrays and their underlying operational principles are closely related to metasurfaces, where the generalized Snell's law allows sub-wavelength manipulation by locally controlling a phase gradient \cite{assouar2018acoustic}. 
This brings about a considerable advantage over resonant metamaterials and Bragg-scattering-based periodic crystals: The wave-manipulating medium is not necessarily the same as that of the wave-carrying one.
Unlike metasurfaces, phased arrays generate and transmit signals rendering them strong candidates for experimental implementation. As a case in point, phase gradients can be conveniently produced via micro-controllers which can be used to create a series of synchronized digital signals with prescribed phase shifts for every element of the array. In order to run the transceivers, the digital signals can then be converted to analog ones using conventional D2A converters.
In this study, we specifically investigate acoustic phased arrays where the wave transceivers are common electromechanical transducers such as piezoelectric patches, speakers, microphones and the like.
While we present in-line acoustic phased arrays as a proof of concept, the physical insights demonstrated here readily extend to higher dimensions which are fairly application-oriented.

In general, phased arrays are capable of operating in both ``transmit" (hereafter denoted by $\mathbb{TX}$) and ``receive" modes (hereafter denoted by $\mathbb{RX}$) \cite{hansen2009phased}. In other words, a phased array can transfuse acoustic waves to an arbitrary direction and ``listen for" acoustic waves incident from an arbitrary direction. By definition, conventional phased arrays exhibit identical radiation patterns between $\mathbb{TX}$ and $\mathbb{RX}$ modes; a direct consequence of the reciprocity principle. Due to time-reversal symmetry, linear time-invariant (LTI) systems exhibit a reciprocal behavior causing transmission between two spatially separated points to remain unchanged following an interchange of the excitation and sensor positions \cite{achenbach2003reciprocity,lorentz1896theorem}. As a result, a given array can only detect waves if they are incident from the same direction that waves can be transmitted to. Although such dual-mode operation aligns well with some real-world applications of phased arrays (e.g., underwater telecommunication of submarines relies on both transmitted and received sound signals for navigation, object detection and obstacle identification), the reciprocity significantly diminishes the scope of their operation. A recent surge of research activity has shown that the intentional breakage of time-reversal symmetry instigates a nonreciprocal behavior in LTI systems which can unlock new opportunities in wave manipulation that were otherwise untapped. Such nonreciprocal systems have the potential to fuel the future of many fields ranging from elasticity \cite{attarzadeh-2018, nassar2017non, chen2020active}, acoustics \cite{li2011tunable,boechler2011bifurcation,yang2015topological, nassar2020nonreciprocity}, and electromagnetics \cite{adam2002ferrite,sounas2017non,shupe1980thermally}, to natural hazard protection and quantum computations.
In optics, there exists a few studies which investigate nonreciprocal wave behavior in antennas \cite{hadad2013one,zang2019nonreciprocal}, space-time phase modulated metasurfaces \cite{ guo2019nonreciprocal,cardin2020surface,zhang2018space, zhang2019breaking}, and real time multi-functional metasurfaces \cite{wang2020theory}. Notable efforts in acoustics attempt to break the symmetry of radiation patterns in transmission and reception as well \cite{maznev2013reciprocity,fleury2015nonreciprocal,devaux2019acoustic}. Means to induce a nonreciprocal behavior in elastoacoustic systems include the exploitation of nonlinearities \cite{moore2018nonreciprocity, popa2014non,liang2010acoustic,liang2009acoustic}, imposing a momentum bias by inducing actual motion \cite{attarzadeh2018, fleury2014sound}, or an artificial momentum bias using space-time modulations \cite{fleury2015subwavelength,attarzadeh2018non,palermo2020surface}. The latter approach is favored due to the insensitivity to wave intensity, the low power consumption, and the feasibility of conducting tests in a finite experimental setup \cite{wang2013optical,attarzadeh2020beam}.

In this report, we propose an \textit{in situ} controllable acoustic phased array with space-time-periodic (STP) phase variation that breaks time-reversal symmetry and enables nonreciprocal phononic transition in both momentum and energy spaces.
By leveraging a dynamic phase modulation provided by a series of phase shifters, the proposed linear array is able to support distinct radiation patterns in transmission and reception that can be tuned independently.
Furthermore, the operational range of the new STP phased array spans multiple directions and frequency channels, simultaneously, rendering it a selective wave-beaming device which can be rapidly and efficiently tuned as desired, as will be detailed and shown here.

\section*{Theoretical Background}
\label{sec:theory}
\begin{figure}[th!]
    \centering
    \includegraphics[width = \linewidth]{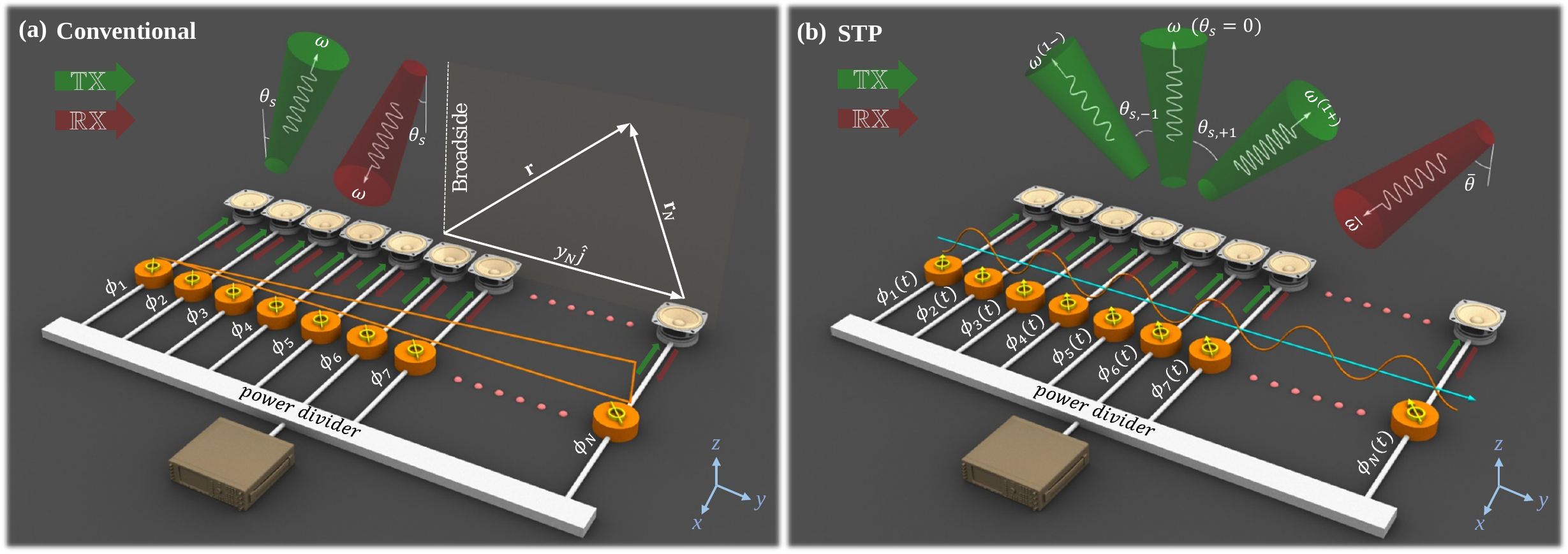}
    \caption{Acoustic phased arrays. (a) Conventional phased array in $\mathbb{TX}$/$\mathbb{RX}$ modes with a static phase gradient. (b) STP phased array in $\mathbb{TX}$/$\mathbb{RX}$ modes with a dynamically changing phase gradient. Green and red colors denote transmitted and incident waves, respectively.}
    \label{fig:phase_array}
    \vspace{-0.5cm}
\end{figure}

A conventional acoustic phased array is illustrated in Fig.~\ref{fig:phase_array}a.
By incorporating a static phase gradient, conventional phased arrays are able to ``transmit" pressure waves that travel in a desired direction in the free space (e.g., $\theta_s$ in Fig.~\ref{fig:phase_array}a as indicated by the green arrow).
Likewise, the array can operate in the ``receive" (listening) mode. Limited by reciprocity, the array will exhibit the strongest gain for waves incident from the same exact $\theta_s$ direction, shown by the red arrow.
The proposed STP linear phased array shown in Fig.~\ref{fig:phase_array}b, however, can defy reciprocity by incorporating a dynamically changing phase angle. In here, we impose a phase angle which follows a prescribed space-time-periodic variation to dynamically vary the signal's phase gradient, contrary to conventional phased arrays with static or quasi-static phase angles.
In the following, we lay out the theoretical framework in transmit ($\mathbb{TX}$) and receive ($\mathbb{RX}$) modes and then describe the breakage of reciprocal symmetry, thereby establishing different and tunable radiation patterns in transmission and reception.

\subsection*{Transmit ($\mathbb{TX}$) Mode}
\label{sec:theory1}
We begin with the STP acoustic phased array in $\mathbb{TX}$ mode.
The array, which is depicted in Fig.~\ref{fig:phase_array}b, comprises $N$ acoustic transducers stacked vertically at spatial intervals equal to $d$.
Each transducer is coupled with a phase shifter which augments the incoming signal with a STP phase angle, $\phi_n(t)$, described by
\begin{equation}\label{eq:STP_phase}
    \phi_n(t) = \kappa_s y_n + \delta \cos(\omega_m t - \kappa_m y_n)
\end{equation}
where $n=1,2,...,N$ is the transducer index, $\kappa_s$ is the static phase gradient (static wavenumber; also present in the conventional phased array), $y_n$ is the vertical position of the $n^{\text{th}}$ transducer along the array, $\delta$ denotes the amplitude of the space-time modulation, $\omega_m$ is the temporal modulation frequency and $\kappa_m$ is the spatial modulation frequency.
Considering a harmonic input signal in the $\mathbb{TX}$ mode, the voltage supplied to the array is $v(t)=V_0 \, \ee^{\ii \omega t}$, where $V_{0}$ is the amplitude, $\omega$ is the temporal frequency, and $\ii=\sqrt{-1}$ is the unit imaginary number.
The coupled STP phase shifters impart an additional phase angle described by Eq.~(\ref{eq:STP_phase}) to the input voltage signal prior to feeding it to the transducers.
Consequently, the supplied voltage to the $n^{\text{th}}$ transducer is $v_n(t) = V_0 \, \ee^{\ii[ \omega t - \phi_n(t)] }$, or
\begin{equation}\label{eq:vol_sig}
    v_n(t) = V_0 \, \ee^{\ii(\omega t-\kappa_s y_n)} \, \ee^{-\ii \delta \cos(\omega_m t - \kappa_m y_n)}
\end{equation}
The exponential term with dynamic phase variation on the right hand-side of Eq.~\prettyref{eq:vol_sig} can be replaced with an infinite series of Bessel functions found by a Jacobi-Anger expansion. 
The result is
\begin{equation}\label{eq:vol-sig2}
    v_n(t) = V_0 \, \ee^{\ii(\omega t - \kappa_s y_n)}  \sum\limits_{q = -\infty}^{\infty} \ii^q \JJ_q(-\delta) \, \ee^{\ii q (\omega_m t - \kappa_m y_n)}
\end{equation}
where $\JJ_q(\bullet)$ denotes the $q^{\text{th}}$-order Bessel function of the first kind.
As can be inferred from Eq.~\prettyref{eq:vol-sig2}, the injected power is theoretically split into an infinite number of harmonic signals. However, by tuning the modulation amplitude $\delta$, a considerable share of energy can be directed to the desired frequency component(s). As such, contributions from second and higher order terms can be reasonably neglected by choosing a relatively small $\delta$. Upon using the identity $J_{-q}=(-1)^{q} J_{+q}$, $v_n(t)$ can be approximated as
\begin{equation} \label{eq:first_order}
    \begin{split}
    v_n(t) &\cong V_0 \JJ_0(\delta) \, \ee^{\ii(\omega t - \kappa_s y_n)} - \ii V_0 \JJ_1(\delta) \left( \ee^{\ii [\omega^{(1+)} t - \kappa_s^{[1+]} y_n]} +  \ee^{\ii[\omega^{(1-)} t - \kappa_s^{[1-]} y_n]} \right)
    \end{split}
\end{equation}
where $(\bullet)^{(q \pm)}$ and $(\bullet)^{[q \pm]}$ are the shorthand notations for a frequency shift of $\pm q\omega_m$ and a wavenumber shift of $\pm q\kappa_m$, respectively.
We note that, only the zeroth and first-order Bessel functions (i.e., $J_0$ and $J_1$) are retained in Eq.~\prettyref{eq:first_order} and the static wavenumber, $\kappa_s$ is carried over to all of the three terms.
Hereafter, the three remaining terms are referred to as the fundamental, up-converted, and down-converted components, respectively from left to right.
The up- and down-converted components are the direct consequences of supplementing the array with STP phase angle and, as a result, disappear once $\delta$ vanishes. We also note the identical coefficient ($ V_0 J_1$) signaling that energy is evenly distributed to both higher and lower frequencies.
Assuming transducers are isotropic and remain in their linear range of operation, i.e., exhibit ideal omnidirectional behavior, the output voltage from each phase-shifter described in Eq.~\prettyref{eq:first_order} is expected to create equivalent acoustic pressure waves of the same frequency content and phase angle.
Hence, it can be shown that the $n^{\text{th}}$ transducer generates three individual pressure waves at a distance $|\mathbf{r}_n|$ away as follows
\begin{equation} \label{eq:press_transmission}
\begin{split}
     p_n (\mathbf{r}_n, t) & \cong \frac{P_0 \JJ_0(\delta)}{|\mathbf{r}_n|} \ee^{\ii(\omega t - \bm{\kappa}.\mathbf{r}_n)} \ee^{-\ii \kappa_s y_n} 
     \\
     &- \frac{\ii P_0 J_1(\delta) }{|\mathbf{r}_n|} \Big ( \ee^{\ii(\omega^{(1+)}t - \bm{\kappa}^{(1+)}.\mathbf{r}_n)} \, \ee^{-\ii \kappa_s^{[1+]} y_n}
     + \ee^{\ii(\omega^{(1-)}t - \bm{\kappa}^{(1-)}.\mathbf{r}_n)} \, \ee^{-\ii \kappa_s^{[1-]} y_n} \Big )
\end{split}
\end{equation}
where $\mathbf{r}_n$ is the spatial position vector with respect to the $n^{\text{th}}$ transducer, $P_0=\mathcal{T}V_{0}$ is the wave amplitude with $\mathcal{T}$ as the transformation coefficient of the transducers.
Here we limit our attention to acoustic transducers with a flat response, which effectively renders $\mathcal{T}$ frequency-independent.
In practice, however, $\mathcal{T}$ may be obtained accurately from transducer's frequency response function.
The wavevectors of the produced acoustic waves are $\bm{\kappa}$, $\bm{\kappa}^{(1+)}$ and $\bm{\kappa}^{(1-)}$, respectively. 
The total acoustic pressure at every spatial point and time is consequently computed by adding the waves generated by individual transducers, i.e., $p_{\text{net}}(\mathbf{r},t) = \sum_{n = 1}^N p_n(\mathbf{r}_n,t)$. Per Fig.~\ref{fig:phase_array}a, the position vector with respect to the origin $\mathbf{r}$ is related to $\mathbf{r}_n$ via $\mathbf{r}=\mathbf{r}_n + y_n \hat{j}$ with $\hat{j}$ being a unit vector in the $y$-direction. As such, the net acoustic pressure becomes
\begin{equation} \label{eq:press_net1}
\begin{split}
     p_{\text{net}} (\mathbf{r}, t)  \cong P_0 \bigg \{& \ee^{\ii(\omega t - \bm{\kappa}.\mathbf{r})} \JJ_0(\delta) \sum_{n = 1}^N \frac{1}{|\mathbf{r}-y_n \hat{j}|} \ee^{\ii y_n(\bm{\kappa}.\hat{j} - \kappa_s)} 
     \\
     &-\ii \ee^{\ii(\omega^{(1+)}t - \bm{\kappa}^{(1+)}.\mathbf{r})} J_1(\delta) \sum_{n = 1}^N \frac{1}{|\mathbf{r}-y_n \hat{j}|} \ee^{\ii y_n [\bm{\kappa}^{(1+)}.\hat{j} - \kappa_s^{[1+]}]}
     \\
     &-\ii \ee^{\ii(\omega^{(1-)}t - \bm{\kappa}^{(1-)}.\mathbf{r})} J_1(\delta) \sum_{n = 1}^N \frac{1}{|\mathbf{r}-y_n \hat{j}|} \ee^{\ii y_n [\bm{\kappa}^{(1-)}.\hat{j} - \kappa_s^{[1-]}]}  \bigg \}
\end{split}
\end{equation}
\begin{figure}[th!]
    \centering
    \includegraphics[width = \linewidth]{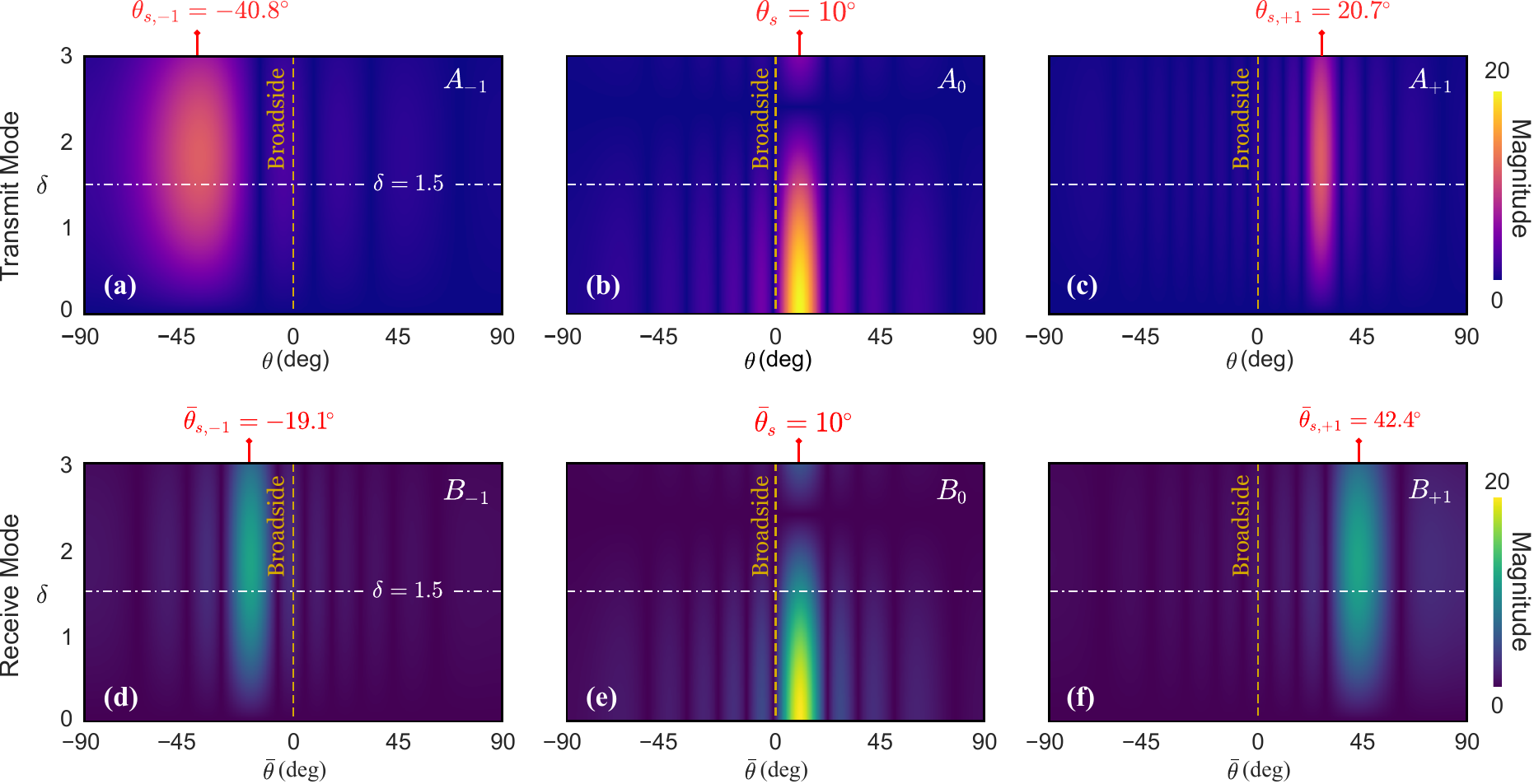}
    \caption{(a-c) Variation of the far-field amplitude coefficients $A_{-1}$, $A_{0}$ and $A_{+1}$ in $\mathbb{TX}$ mode as a function of $\delta$ and $\theta$. The vertical red arrows indicate the principal transmission directions $\theta_{s, -1} = -40.8^\circ$, $\theta_{s} = 10^\circ$, and $\theta_{s, +1} = 20.7^\circ$. (d-f) Variation of the far-field amplitude coefficients $B_{-1}$, $B_{0}$ and $B_{+1}$ in $\mathbb{RX}$ mode as a function of $\delta$ and $\theta$. The vertical red arrows indicate the principal listening directions $\bar{\theta}_{s, -1} = -19.1^\circ$, $\bar{\theta}_{s} = 10^\circ$, and $\bar{\theta}_{s, +1} = 42.4^\circ$. Parameters used are as follows: $\omega/2\pi = 1000$ Hz, $\omega_m/2\pi =500$ Hz, $\kappa_s = 1.0125\pi$ rad/m, $\kappa_m = 2.9154\pi$ rad/m, and $N=20$. The white dotted lines indicate $\delta=1.5$.}
    \label{fig:As}
\end{figure}

In the far field, the magnitude of $|\mathbf{r}-y_n \hat{j}|$ can be approximated as $|\mathbf{r}|$, which reduces Eq.~\prettyref{eq:press_net1} to
\begin{equation} \label{eq:press_net2}
     p (\mathbf{r}, t)_{\text{net}} \cong \frac{P_0}{|\mathbf{r}|}\left \{ A_{0} \, \ee^{\ii(\omega t - \bm{\kappa}.\mathbf{r})} 
     -\ii A_{+1} \, \ee^{\ii(\omega^{(1+)}t - \bm{\kappa}^{(1+)}.\mathbf{r})} -\ii A_{-1} \, \ee^{\ii(\omega^{(1-)}t - \bm{\kappa}^{(1-)}.\mathbf{r})}  \right \}
\end{equation}

In Eq.~\prettyref{eq:press_net2}, we have three dominant spherical waves, fundamental, up- and down-converted, each of them propagating at different frequencies and wavenumbers. The up- and down-conversions in the second and third terms are reminiscent of the phononic transition in both energy and momentum spaces as a result of the space-time periodicity.
The coefficients $A_0$, $A_{-1}$, and $A_{+1}$ of the three waves are dependent on the modulation amplitude $\delta$ and the propagation direction $\theta$, as follows:
\begin{subequations} \label{eq:tra_coeff}
\begin{align}
    A_{0}(\delta,\theta)=& \JJ_0(\delta) \sum_{n = 1}^N \ee^{-\ii (\kappa_s - \kappa \sin \theta) y_n}\label{eq:tra_coeffa}
    \\
    A_{+1}(\delta,\theta)=& J_1(\delta) \sum_{n = 1}^N \ee^{-\ii [ \kappa_s^{[1+]}-\kappa^{(1+)} \sin \theta]y_n}\label{eq:tra_coeffb}
    \\
    A_{-1}(\delta,\theta)=& J_1(\delta) \sum_{n = 1}^N \ee^{-\ii [\kappa_s^{[1-]}-\kappa^{(1-)} \sin \theta] y_n}\label{eq:tra_coeffc}
\end{align}
\end{subequations}

In writing Eq.~\prettyref{eq:tra_coeff}, we considered an arbitrary $\theta$ for waves, which leads to wavevectors as $\bm{\kappa}^{(q\pm)}=\kappa^{(q\pm)}(\cos \theta \hat{i} + \sin \theta \hat{j})$ for $q=0,1$ with wavenumbers given by $\kappa^{(q\pm)}=\frac{\omega^{(q\pm)}}{c}$ and $c$ as the speed of sound in air.
Since $J_0(0)=1$ and $J_1(0)=0$, it can be verified that both the $A_{+1}$ and $A_{-1}$ terms vanish as soon as $\delta=0$ and only the fundamental wave component remains, which brings us back to the conventional phased array.
The variation of the three components with respect to $\theta$ and $\delta$ are more clearly illustrated in Fig.~\ref{fig:As}a-c, where the color intensity indicates the strength of each wave component in different directions as $\delta$ varies on the $y$-axis.
We observe that the STP phased array exhibits three independent principal $\mathbb{TX}$ channels, each operating at a different frequency (namely $\omega$, $\omega^{(1+)}$ and $\omega^{(1-)}$) and is capable of transmitting waves in different non-trivial directions.
A closer inspection of Eqs.~\prettyref{eq:press_net2} and \prettyref{eq:tra_coeff} also reveals how these three principal directions can be calculated.
For example, from Eq.~\prettyref{eq:tra_coeffa}, we find that the coefficient of the fundamental wave component $A_0$ is dominant in a direction that nullifies the argument of its exponential term for any given $n$. 
This implies that the fundamental wave component predominantly propagates along the $\theta = \sin ^{-1} (\frac{c \kappa_s}{\omega})$ direction. 
While the previous is also a feature of conventional arrays, by setting the argument of the exponential term in Eq.~\prettyref{eq:tra_coeffb} equal to zero, the up-converted wave mode now travels primarily in a direction that is given by $\sin \theta = \frac{c \kappa_s^{[1+]}}{\omega^{(1+)}}$, which maximizes magnitude of $A_{+1}$.
The same feature extends to the coefficient of the down-converted wave $A_{-1}$ and we get propagation in a direction that satisfies $\sin \theta =\frac{c \kappa_s^{[1-]}}{\omega^{(1-)}}$.
The two aforementioned propagation directions in the $\mathbb{TX}$ mode are absent in conventional systems. These three principal transmission directions can be visualized in the green arrows of Fig.~\ref{fig:phase_array}b and are denoted by $\theta_s$, $\theta_{s,+1}$ and $\theta_{s,-1}$, defined as
\begin{align} \label{eq:thetas}
    \theta_{s,\pm q}=&\sin^{-1} \left (\frac{c \kappa_s^{[q\pm]}}{\omega^{(q\pm)}} \right)
    \quad \quad \text{for} \quad \quad q=0,1
\end{align}
\begin{figure}[ht!]
    \centering
    \includegraphics[width = 0.82\linewidth]{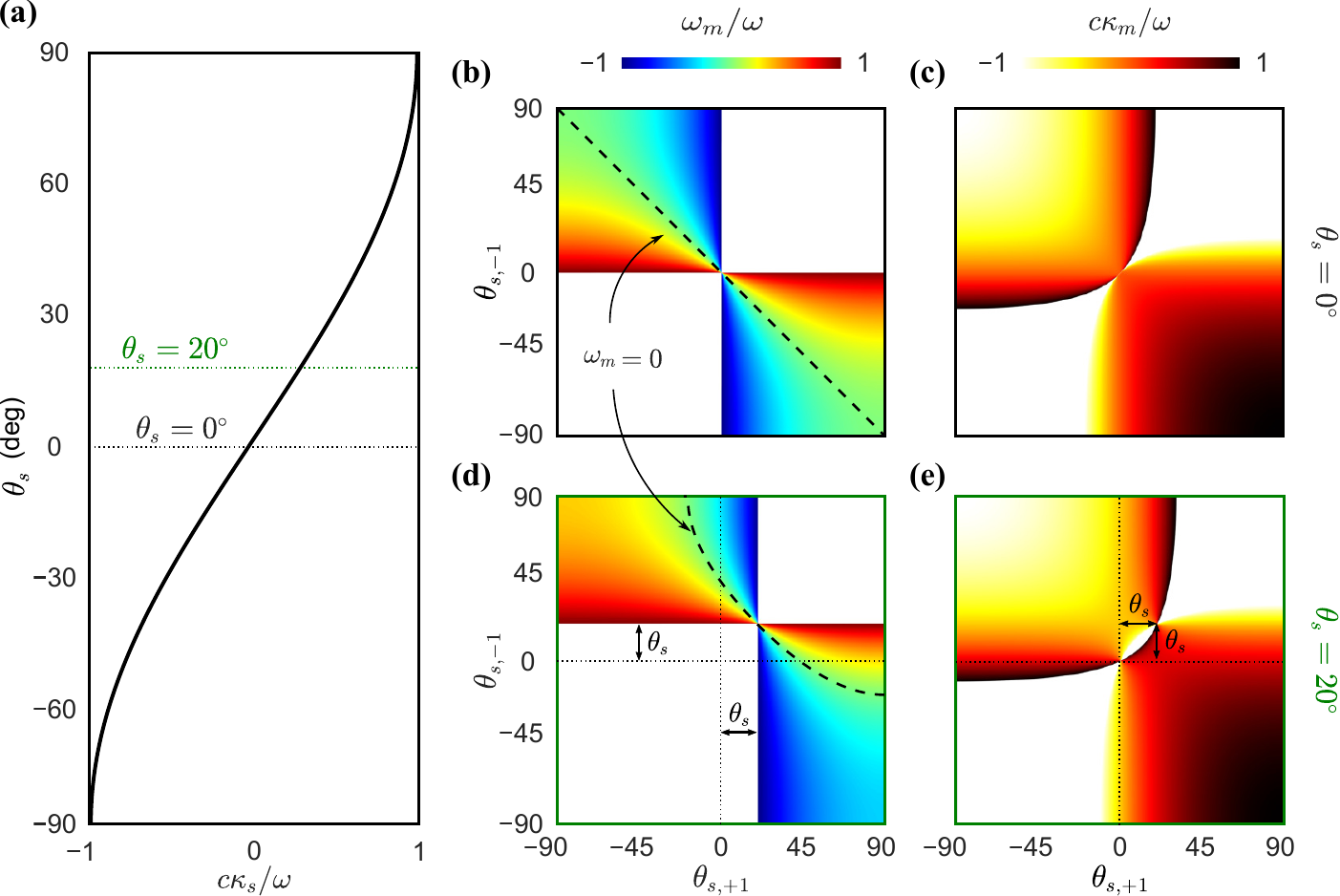}
    \caption{(a) Variation of the design parameter $c\kappa_s/\omega$ with respect to $\theta_s$. (b-c) The design parameters $\omega_m/\omega$ and $c\kappa_s/\omega$ as functions of $\theta_{s, -1}$ and $\theta_{s. +1}$ for a fixed $\theta_s = 0$. The dashed line indicates $\omega_m = 0$. (d-e) The design parameters $\omega_m/\omega$ and $c\kappa_s/\omega$ as functions of $\theta_{s, -1}$ and $\theta_{s. +1}$ for a fixed $\theta_s = 20$. The dashed curve indicates $\omega_m = 0$.}
    \label{fig:tra_parameters}
\end{figure}

Instead of using the parameters of the STP array to find these directions, we can alternatively solve the three equations in Eq.~\prettyref{eq:thetas} with $q=0$ and $1$. Solving for the three tunable parameters $\kappa_m$, $\kappa_s$, and $\omega_m$ as a function of known values of $c$, $\omega$, $\theta_s$, $\theta_{s,+1}$, and $\theta_{s,-1}$ yields a unique set of parameters that enable a desired performance. The outcome is
\begin{subequations} \label{eq:parameters}
\begin{align}
     \frac{c \kappa_s}{\omega}=&  \sin \theta_{s}
    \\
     \frac{\omega_m}{\omega} = & \frac{2 \sin \theta_{s}}{\sin \theta_{s,+1} - \sin \theta_{s,-1}} - \frac{\sin \theta_{s,+1}+\sin \theta_{s,-1}}{\sin \theta_{s,+1} - \sin \theta_{s,-1}}
     \\
     \frac{c \kappa_m}{\omega} = & \sin \theta_{s} \frac{ \sin \theta_{s,-1} + \sin \theta_{s,+1}}{\sin \theta_{s,+1} -\sin \theta_{s,-1}} - \sin \theta_{s,+1} \frac{ 2 \sin \theta_{s,-1}}{\sin \theta_{s,+1} -\sin \theta_{s,-1}}
\end{align}
\end{subequations}

The previous approach is particularly useful in the design and operation of acoustic phased arrays in a scenario where transmitting signals in various prescribed directions with \textit{in situ} tunability is highly desirable. Figure~\ref{fig:tra_parameters} shows the variation of the left hand side of Eqs.~\prettyref{eq:parameters} as a function of $\theta_{s,\pm 1}$ for $\theta_s=0$ and $\theta_s=20^\circ$.
It can be observed that while $\kappa_s$ monotonically decreases with $\theta_s$, the response of the other parameters of the STP phased array are more complex. 
For instance, in Fig.~\ref{fig:tra_parameters}b, we deduce that waves can be transmitted in perfect symmetric directions (i.e., $\theta_{s,+1}=-\theta_{s,-1}$), by setting $\kappa_s=0$ and $\omega_m=0$ as indicated by the dashed line.
In Figs.~\ref{fig:tra_parameters}d and e, the shift by the $\theta_s$ value is clearly apparent compared to the same parameters in b and c (where $\theta_s=0$). To conclude, the framework shown here depicts a non-trivial and unprecedented level of control over both the direction and frequency (channel) of the transmitted wave beams in the STP array, which solely emerges as a consequence of the imposed space-time modulation. The simultaneous transition in momentum (wavenumber) and energy (frequency) spaces brought about by such modulation open up the possibility of multi-direction and multi-channel wave-beaming effects, respectively.

\subsection*{Receive ($\mathbb{RX}$) Mode}
\label{sec:theory2}
In the $\mathbb{RX}$ mode, incident acoustic waves are first converted back into electrical signal by the transducers and are then sent back through the STP phase shifters to be collected at the output channel, thus enabling detection of objects which reflect waves or sources that emit waves.
Let us consider a plane-wave acoustic beam that is incident upon the STP phased array from an arbitrary direction $\bar{\theta}$ measured from broadside with a temporal frequency $\bar{\omega}$ and a wavenumber $\bar{\kappa}=\frac{\bar{\omega}}{c}$, as illustrated in Fig.~\ref{fig:phase_array}b with a red arrow.
Owing to the spatial spacing between the array receivers, the beam experiences a time delay in reaching farther transducers. Specifically, a phase shift of $\bar{\kappa} y_n \sin \bar{\theta}$ is induced at the $n^{\text{th}}$ transducer. Consequently, the voltage generated by the $n^{th}$ transducer can be given by $\bar{V}_0 \ee^{\ii(\bar{\omega} t + \bar{\kappa}y_n \sin \bar{\theta})}$ with $\bar{V}_{0}$ being the voltage amplitude.
In the previous, the transducers were implicitly kept linear, isotropic, and exhibit a flat frequency response---similar to the $\mathbb{TX}$ mode.
After passing through the dynamic STP phase shifter, the output voltage signal collected at the $n^{th}$ transducer becomes
\begin{equation}\label{eq:vol-gen-rec}
    \bar{v}_n(t) = \bar{V}_0 \ee^{\ii(\bar{\omega} t + \bar{\kappa} y_n \sin{\bar{\theta}})} \, \text{e}^{-\text{i}[\kappa_s y_n + \delta \cos(\omega_m t - \kappa_m y_n)]}
\end{equation}
which, using the Jacobi-Anger expansion one more time, gives
\begin{equation}\label{eq:vol-gen-rec2}
    \bar{v}_n(t) = \bar{V}_0 \ee^{\ii[\bar{\omega} t - (\kappa_s - \bar{\kappa}\sin \bar{\theta}) y_n]}  \sum\limits_{q = -\infty}^{\infty} \ii^q \JJ_q(-\delta) \ee^{\ii q (\omega_m t - \kappa_m y_n)}\\
\end{equation}
Assuming a sufficiently small modulation amplitude $\delta$, we only retain contributions from the $\JJ_0$ and $\JJ_{\pm 1}$ terms, which reduces Eq.~\prettyref{eq:vol-gen-rec2} to
\begin{equation} \label{eq:vol-rec-decouple}
\begin{split}
\bar{v}_n(t) \cong & \bar{V}_0 \JJ_0(\delta)\ee^{\ii[\bar{\omega} t - (\kappa_s - \bar{\kappa} \sin \bar{\theta})y_n]}
\\
&- \ii\bar{V}_0 \JJ_1(\delta) \left (\ee^{\ii[\bar{\omega} ^{(1+)}t - (\kappa_s^{[1+]} - \bar{\kappa} \sin \bar{\theta})y_n]} + \ee^{\ii[\bar{\omega} ^{(1-)}t - (\kappa_s^{[1-]} - \bar{\kappa} \sin \bar{\theta} )y_n]} \right)
\end{split}
\end{equation}
which follows the same short-hand notation introduced earlier in the $\mathbb{TX}$ mode.
As per Fig.~\ref{fig:phase_array}b, the output channel at the listening port receives a summation of all the $n$ signals (i.e., $\bar{v}(t)=\sum_{n=1}^N \bar{v}_n(t)$), which after a few simplifications can be broken down into three signal components:
\begin{equation} \label{eq:main_receive}
    \bar{v}(t)\cong \bar{V}_0 \left\{ B_0 \ee^{\ii\bar{\omega} t} 
    -\ii B_{+1}  \ee^{\ii\bar{\omega} ^{(1+)}t}
    -\ii B_{-1}  \ee^{\ii\bar{\omega} ^{(1-)}t} \right \}
\end{equation}
where the amplitude of each is given by
\begin{subequations} \label{eq:main_receive2}
\begin{align}
    B_0 (\delta,\bar{\theta})=& \JJ_0 (\delta) \sum_{n=1}^N \ee^{-\ii (\kappa_s - \bar{\kappa} \sin \bar{\theta})y_n}
    \\
    B_{+1}(\delta,\bar{\theta})=& \JJ_1(\delta) \sum_{n=1}^N \ee^{- \ii (\kappa_s^{[1+]} - \bar{\kappa} \sin \bar{\theta})y_n}
    \\
    B_{-1}(\delta,\bar{\theta})=& \JJ_1(\delta) \sum_{n=1}^N \ee^{- \ii (\kappa_s^{[1-]}  - \bar{\kappa} \sin \bar{\theta})y_n}
\end{align}
\end{subequations}

Figure~\ref{fig:As}d-f reveals the variation of the magnitudes of $B_0$, $B_{+1}$ and $B_{-1}$ as a function of the modulation amplitude $\delta$ and the incident direction $\bar{\theta}$. Unlike conventional phased arrays with a single principal listening direction, three dominant directions emerge in the STP phased array and are apparent in the figure as predicted. Upon inspection, Eq.~\prettyref{eq:main_receive} is effectively the $\mathbb{RX}$-equivalent of Eq.~\prettyref{eq:press_net2} in the $\mathbb{TX}$ mode, and can therefore be used to identify the three principal listening directions. Eqs.~\prettyref{eq:main_receive} and \prettyref{eq:main_receive2} show that the STP array has three independent listening channels, each operating at a different frequency, namely $\bar{\omega}^{0}$, $\bar{\omega}^{(1+)}$, and $\bar{\omega}^{(1-)}$.
Following the same reasoning given in the $\mathbb{TX}$ mode, if an incoming signal is incident from a $\bar{\theta}$ that satisfies $\sin \bar{\theta}=\frac{c \kappa_s}{\bar{\omega}}$, then $B_0$ becomes dominant and the fundamental signal component with $\ee^{\ii\bar{\omega} t}$ will be most efficiently detected. While the previous is also a feature of conventional arrays, what is unique here is that if $\sin \bar{\theta}=\frac{c \kappa_s^{[1+]}}{\bar{\omega}}$, then $B_{+1}$ becomes dominant and the up-converted signal component can be detected along the $\bar{\theta}$ direction. 
The same applies to the down-converted component.
Finally, and as predicted, both $B_{-1}$ and $B_{+1}$ disappear by setting $\delta$ equal to zero. We refer to the three principal listening directions as $\bar{\theta}_s$, $\bar{\theta}_{s,+1}$, and $\bar{\theta}_{s,-1}$, and summarize them as follows
\begin{align} \label{eq:barthetas}
    \bar{\theta}_{s,\pm q} =\sin^{-1} \left ( \frac{c\kappa_s^{[q\pm]}}{\bar{\omega}}\right) \quad \quad \text{for} \quad q=0,1
\end{align}

\begin{figure}[htbp!]
    \centering
    \includegraphics[width = \linewidth]{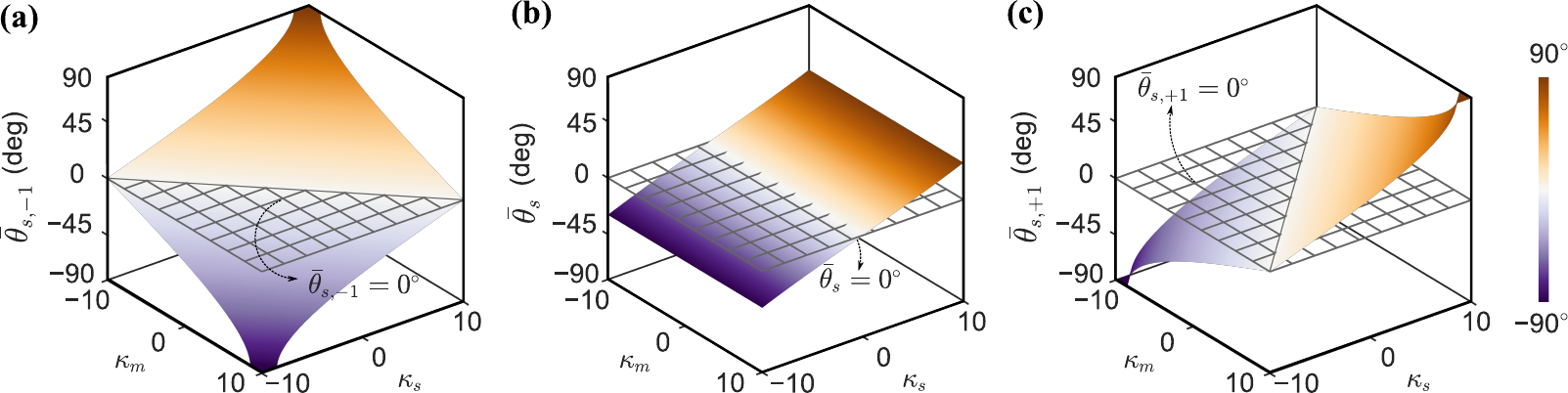}
    \caption{Effect of changing $\kappa_m$ and $\kappa_s$ on the three principal listening directions: (a) $\bar{\theta}_{s, -1}$, (b) $\bar{\theta}_{s}$, and (c) $\bar{\theta}_{s, +1}$.}
    \label{fig:barthetas}
\end{figure}

Fig.~\ref{fig:barthetas} shows the sensitivity of these three angles to the phased array parameters.
Interestingly, all the angles are independent of $\omega_m$ and are shown here as functions of $\kappa_s$ and $\kappa_m$.
We note that all of the three principal listening directions in Eq.~\prettyref{eq:barthetas} are \textit{in situ} tunable and can be turned towards three different spatial points. 
Furthermore, the listening directions can operate simultaneously without interference and are different than the principal transmission directions given by $\theta_{s,\pm q}$ in Eq.~\prettyref{eq:thetas}.
The former is evidence of asymmetry between radiation patterns in $\mathbb{TX}$ and $\mathbb{RX}$ modes, which will be further discussed in detail in the following subsection, while the latter exemplifies the scanning capabilities of the STP array on top of the multi-directional wave beaming demonstrated earlier in the $\mathbb{TX}$ mode.

\subsection*{Nonreciprocal Behavior}
Reciprocity is an integral hallmark feature of linear time-invariant systems. In a reciprocal system, transmission between any two arbitrary points remains unchanged if the actuator and sensor locations are interchanged. To demonstrate nonreciprocity in the STP phased array, we excite it with a simple harmonic input $v(t)=V_0 \, \ee^{\ii \omega t}$ in the $\mathbb{TX}$ mode. 
Per Eq.~\prettyref{eq:press_net2}, we anticipate acoustic waves to propagate in three distinct channels (fundamental, up-converted, and down-converted), each having a unique frequency ($\omega$, $\omega^{(1+)}$, and $\omega^{(1-)}$) and direction ($\theta_{s}$, $\theta_{s, +1}$, and $\theta_{s, -1}$). In the $\mathbb{RX}$ mode, we consider the time-reversed waves, i.e., the same three wave components traveling in reversed directions and incident upon the array. 
The comparison between these two modes can reveal breakage of reciprocal symmetry in the STP phased array.

Starting with the fundamental channel, which carries a wave of frequency $\bar{\omega}=\omega$ incident from $\bar{\theta}=\theta_{s}$, it can be shown from Eq.~\prettyref{eq:main_receive} that $\bar{v}(t)=\bar{V}_0 B_{0}  \ee^{\ii \omega t}$ is the dominantly received signal. In this case, $\bar{v}(t)$ has the same frequency content as that of $v(t)$, rendering the fundamental channel reciprocal. The up-converted channel hosts a wave of frequency $\bar{\omega}=\omega^{(1+)}$ incident from $\bar{\theta}=\theta_{s,+1}$.
The same equation implies that the array will dominantly up-convert this signal, leading to a received signal of $\bar{v}(t)=-\ii \bar{V}_0 B^{(1+)}_{+1}  \ee^{\ii \omega^{(2+)}t}$. 
This double up-conversion in $\bar{v}(t)$ compared to $v(t)$ is in itself evidence of nonreciprocal behavior within the array. A similar observation can be made following an analysis of the down-converted channel. In this case, a wave of frequency of $\bar{\omega}=\omega^{(1-)}$ incident upon the array from $\bar{\theta}=\theta_{s,-1}$ will be dominantly down-converted resulting in $\bar{v}(t) = -\ii \bar{V}_0 B^{(1-)}_{-1}  \ee^{\ii \omega^{(2-)}t}$. Such nonreciprocity materializes in the up- and down-converted channels only while the fundamental channel remains reciprocal. This is further confirmed using a derivation of the Scattering matrix, which is detailed in the Supplementary Information.

\begin{figure}[th!]
    \centering
    \includegraphics[width = \linewidth]{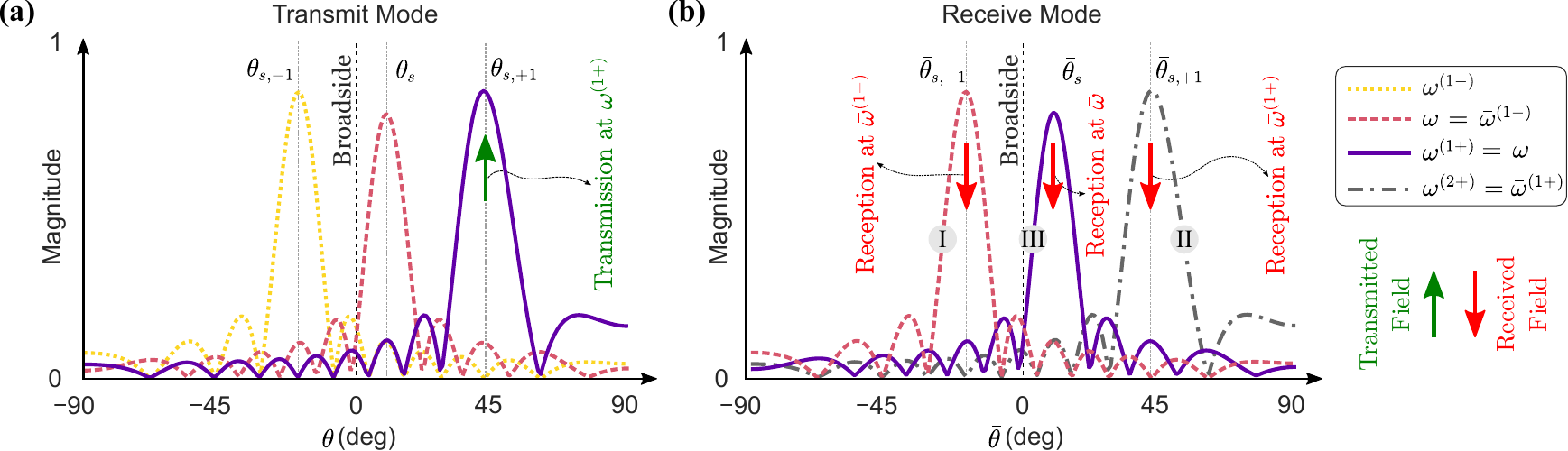}
    \caption{(a) Transmission pattern of the STP array in $\mathbb{TX}$ mode. The up-converted transmission channel is represented with the rightmost curve and the green arrow. (b) Listening pattern of the STP array in $\mathbb{RX}$ mode for an incident wave with a frequency of $\bar{\omega}=\omega^{(1+)}$. Down-converted, up-converted, and fundamental listening channels are denoted by the red arrows and are marked (I) through (III). These will be detected if the incident direction is $\bar{\theta}_{s, -1}$, $\bar{\theta}_{s, +1}$, or $\bar{\theta}_{s}$, respectively. Upon comparing the three listening patterns (I)-(III) with the transmission pattern marked with the green arrow in (a), different types of nonreciprocity (momentum, frequency, and double) are shown to have materialized. Parameters used are as follows: $\delta=1.5$, $\kappa_s = 1.0125\pi$ rad/m, $\kappa_m = 2.9154\pi$ rad/m, and $\omega_m/\omega=0.01$.}
    \label{fig:nonrec1}
\end{figure}

To take a closer look at the various ways in which a nonreciprocal behavior manifests itself in the STP phased array, we examine the $\mathbb{TX}$ radiation pattern depicted in Fig.~\ref{fig:nonrec1}a. The up-converted wave component (solid line) is generated using an input voltage with a frequency $\omega$ and propagates along the $\theta_{s,+1}$ direction. Upon sending back a wave with the same frequency (i.e., $\bar{\omega}=\omega^{(1+)}$), three different scenarios can emerge in the $\mathbb{RX}$ mode, as shown in Fig.~\ref{fig:nonrec1}b:
(I) A down-conversion takes place resulting in an output signal of frequency $\bar{\omega}^{(1-)}=\omega$ and a principal listening direction $\bar{\theta}_{s,-1}$. In this case, the input voltage signal (in $\mathbb{TX}$) and the output one (in $\mathbb{RX}$) share the same frequency but maintain different principal transmission and listening directions -- Reciprocity is broken in the momentum space. (II) An up-conversion takes place resulting in an output signal of frequency $\bar{\omega}^{(1+)}=\omega^{(2+)}$ and a principal listening direction $\bar{\theta}_{s,+1}$. Here, the principal transmission and listening directions are identical, but the frequencies of the input and output voltage signals become different -- Reciprocity is broken in the frequency space.
(III) No conversion takes place resulting in an output signal of frequency $\bar{\omega}=\omega^{(1+)}$ and a principal listening direction $\bar{\theta}_s$. 
In other words, the input and output voltage signals have different frequencies along with different directions associated with the principal transmission and reception directions -- Reciprocity is broken in both frequency and momentum spaces. Following a similar analysis, the same conclusions can be drawn if the down-converted wave component was considered in the first place.

In an intuitive sense, identical acoustic radiation patterns in $\mathbb{TX}$ and $\mathbb{RX}$ modes are also an embodiment of the reciprocity principle; a feature which conventional acoustic phased arrays are bound to exhibit \cite{balanis2016antenna}. However, an STP phased array does not necessarily adhere to this criterion. As a reflection of this, principal transmission and listening directions no longer coincide once a temporal modulation kicks in. This hypothesis can be easily tested out by inspecting the principal directions of each mode described by Eqs.~\prettyref{eq:thetas} and \prettyref{eq:barthetas}. For a relatively slow temporal modulation of $\omega_m/\omega \leq 0.1$, a binomial approximation can be used to obtain
\begin{align} \label{eq:9approx}
    \sin {\theta}_{s,\pm q} \cong \frac{c(\kappa_s\pm q \kappa_m)}{\omega}(1\mp q \frac{\omega_m}{\omega})\quad \quad \text{for} \quad q=0,1
\end{align}
from Eq.~\prettyref{eq:thetas} for the $\mathbb{TX}$ mode.
Without loss of generality, consider a specific case where $\bar{\omega}=\omega$.
As a result, the difference between the sines of the principal listening and transmission directions can be approximated as
\begin{align} \label{eq:diff}
    \Delta_{s,\pm q} \cong \mp q \, \frac{c(\kappa_s\pm q \kappa_m)}{\omega}( \frac{\omega_m}{\omega})\quad \quad \text{for} \quad q=0,1
\end{align}

\begin{figure}[h!]
    \centering
    \includegraphics[width = \linewidth]{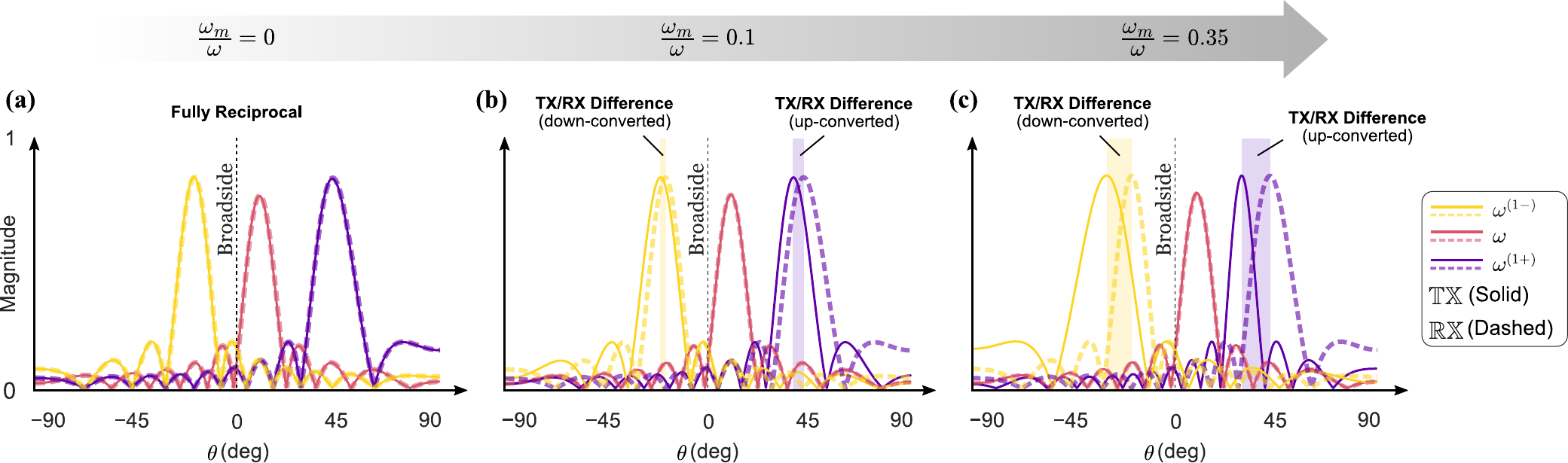}
    \caption{Nonreciprocity in the STP phased arrays demonstrated by a comparison between acoustic radiation patterns in $\mathbb{TX}$ and $\mathbb{RX}$ modes for increasing temporal modulations: (a) $\omega_m/\omega = 0$ (no modulation), (b) $\omega_m/\omega = 0.1$, and (c) $\omega_m/\omega = 0.35$. The parameters used are as follows: $\omega = \bar{\omega} = 1000$ Hz, $\kappa_s = 1.0125 \pi$, $\kappa_m = 2.9154\pi$ and $\delta = 1.5$.}
    \label{fig:non-rec-rad}
\end{figure}

Although $\Delta_{s,\pm q}$ vanishes for $q=0$ (corresponding to the fundamental component), it takes a nonzero value for the up- and down-converted wave components.
It is also noted that such difference between radiation patterns in transmission and reception becomes stronger as the temporal modulation becomes faster, a behavior which is shown in Fig.~\ref{fig:non-rec-rad} which displays $\mathbb{TX}$ (solid) and $\mathbb{RX}$ radiation (dashed) patterns for increasing values of $\omega_m/\omega$. Fig.~\ref{fig:non-rec-rad}a represents the non-modulated system where the difference between $\theta_{s, \pm q}$ and $\bar{\theta}_{s, \pm q}$ disappears as expected from a reciprocal array. Figs.~\ref{fig:non-rec-rad}b and c correspond to $\omega_m/\omega = 0.1$ and $0.35$ and show increasing differences between the solid and dashed lines, respectively. Finally, we note that for $\omega_m/\omega > 0.1 $, the conclusions drawn here remain valid although the binomial expansion may no longer be accurate.

\section*{Results and Discussions}
\label{sec:fullwave}

We report on the transient performance of the STP phased array by using a semi-analytical in-house algorithm that does not incorporate far-field approximations.
We consider 20 acoustic transducers arranged linearly along the $y$-axis. The transducers are separated by a quarter wavelength distance $\lambda/4$ where $\lambda$ is the wavelength of the fundamental component and are centered at the origin. The simulations are carried out up to $1$ second with a sampling frequency of $4000$ Hz on a 2-dimensional domain of size $7 \times 14$ m$^2$, which is discretized using a grid of $251 \times 501$ spatial points.
Conventionally, the acoustic transducers are modeled as dipoles rather than isotropic transducers.
Therefore to mimic practical conditions, we account for the directional behavior of the dipoles by integrating a $Q_n$ coefficient in the acoustic pressure waves of each transducer.
As such, the generated acoustic waves from the $n^{\text{th}}$ transducer---earlier given in Eq.~\prettyref{eq:press_transmission}---become dependent on $\theta$ as follows
\begin{equation} \label{eq:dir_press_transmission}
\begin{split}
     p_n (\mathbf{r}_n,\theta, t)&  \cong \frac{\JJ_0(\delta) }{|\mathbf{r}_n|} Q_n \, \ee^{\ii(\omega t - \bm{\kappa}.\mathbf{r}_n)} \ee^{-\ii \kappa_s y_n} 
     \\
     &- \frac{\ii J_1(\delta) }{|\mathbf{r}_n|} \Big ( Q_n^{(1+)} \, \ee^{\ii(\omega^{(1+)}t - \bm{\kappa}^{(1+)}.\mathbf{r}_n)} \, \ee^{-\ii \kappa_s^{[1+]} y_n}
     + Q_n^{(1-)} \, \ee^{\ii(\omega^{(1-)}t - \bm{\kappa}^{(1-)}.\mathbf{r}_n)} \, \ee^{-\ii \kappa_s^{[1-]} y_n} \Big )
\end{split}
\end{equation}
where $Q_n^{(q\pm)}$ is a function of the wavenumber, directivity coefficient $D_n$ and incorporates other dipole parameters.
For each source, the dipole directivity coefficient is defined  as $D_n=D(\varphi_n,\kappa)=\sin(\frac{1}{2}\kappa b\cos\varphi_n)$, where $b$ is the diameter of the dipole, $\kappa$ is the wavenumber and $\varphi_n$ is the locally measured polar angle \cite{kinsler2009}.
Regardless of the wavenumber, we see that $D=0$ for $\varphi=(2k+1)\frac{\pi}{2}$ with an integer $k$---since commercially available speakers are incapable of propagating waves in their respective end-fire direction.
Based on the geometry of the phased array shown in Fig.~\ref{fig:phase_array}a, we can verify that $\varphi_n$ is related to $\theta$ through $\varphi_{n}=\tan^{-1}(\tan \theta - \frac{y_n}{|\mathbf{r}|} \sec \theta )$ for the $n^{\text{th}}$ transducer. Including the far-field approximation $y_n \ll |\mathbf{r}|$, this relation simplifies to $\varphi_n=\theta$ for all $n=1,2,..,N$.

\vspace{0.3cm}

\begin{figure}[th!]
    \centering
    \includegraphics[width = 0.75\linewidth]{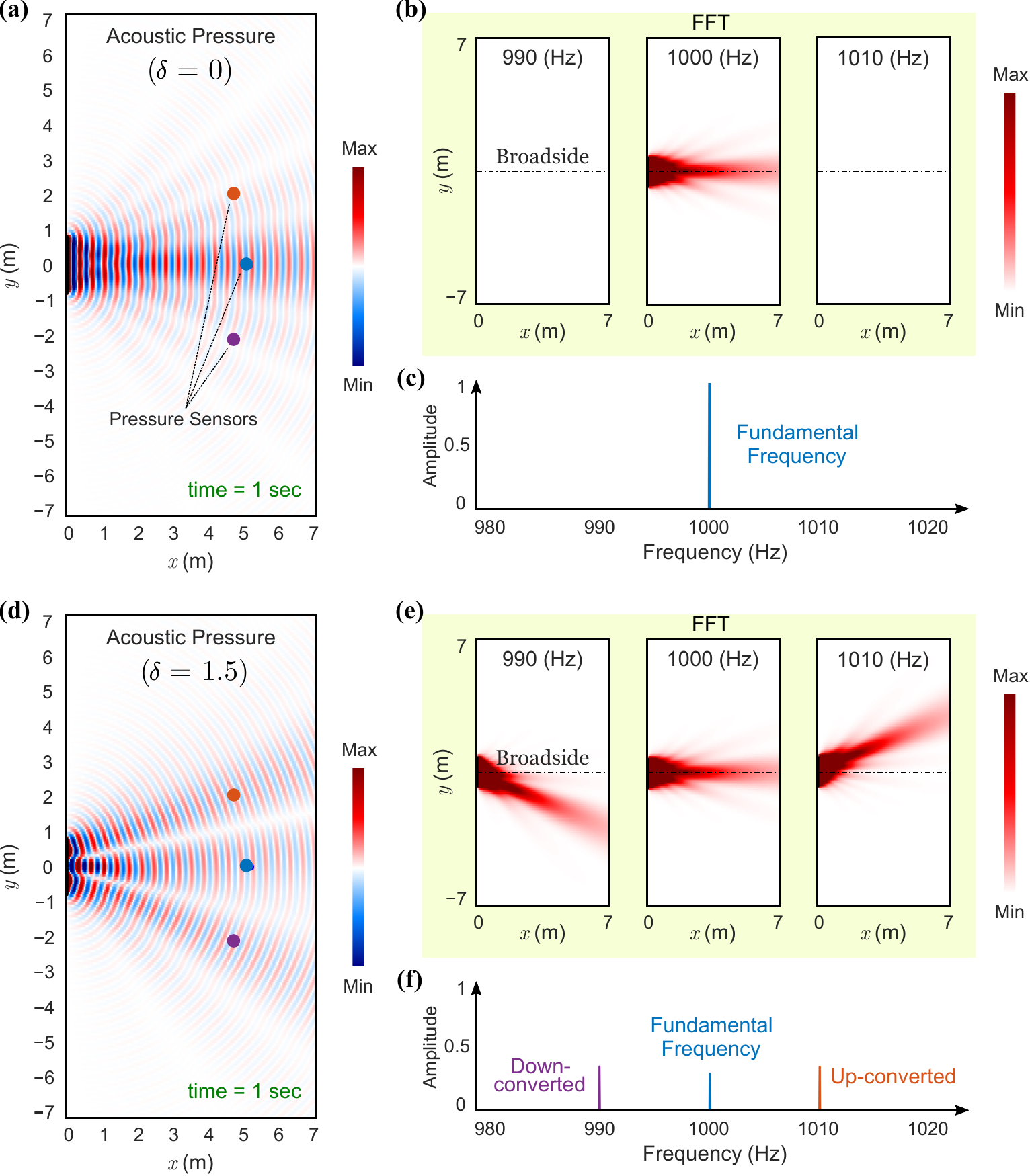}
    \caption{Time transient acoustic pressure field at time $= 1$ s for (a) $\delta = 0$ and (d) $\delta = 1.5$. The fundamental frequency of the supplied voltage is $\omega/2\pi = 1000$ Hz and $\kappa_s$ is set equal to zero. The spatial and temporal modulation frequencies are $\kappa_m = 2.3344\pi$ and $\omega_m/2\pi = 10$ Hz, respectively. In the top panel, the fundamental wave component propagates along the broadside direction only. In (b), the FFT amplitudes of the entire considered space are shown for three frequencies: $990$ Hz, $1000$ Hz, and $1010$ Hz for $\delta = 0$. In (c), the FFTs of pressure waves at the three sensor locations marked with red ($23^\circ$), blue ($0^\circ$), and purple ($-24^\circ$) are shown for $\delta = 0$. In the bottom panel, the fundamental wave component propagates along the broadside, are up-converted at $23^\circ$, as well as down-converted at $-24^\circ$ for $\delta = 1.5$. In (e), the FFT amplitudes of the entire considered space are shown for three frequencies: $990$ Hz, $1000$ Hz, and $1010$ Hz for $\delta = 1.5$. In (f), the FFTs of pressure waves at the three sensor locations marked with red ($23^\circ$), blue ($0^\circ$), and purple ($-24^\circ$) are shown for $\delta = 1.5$.} 
    \label{fig:matlab}
\end{figure}

\begin{figure}[th!]
\centering
\includegraphics[width = \linewidth]{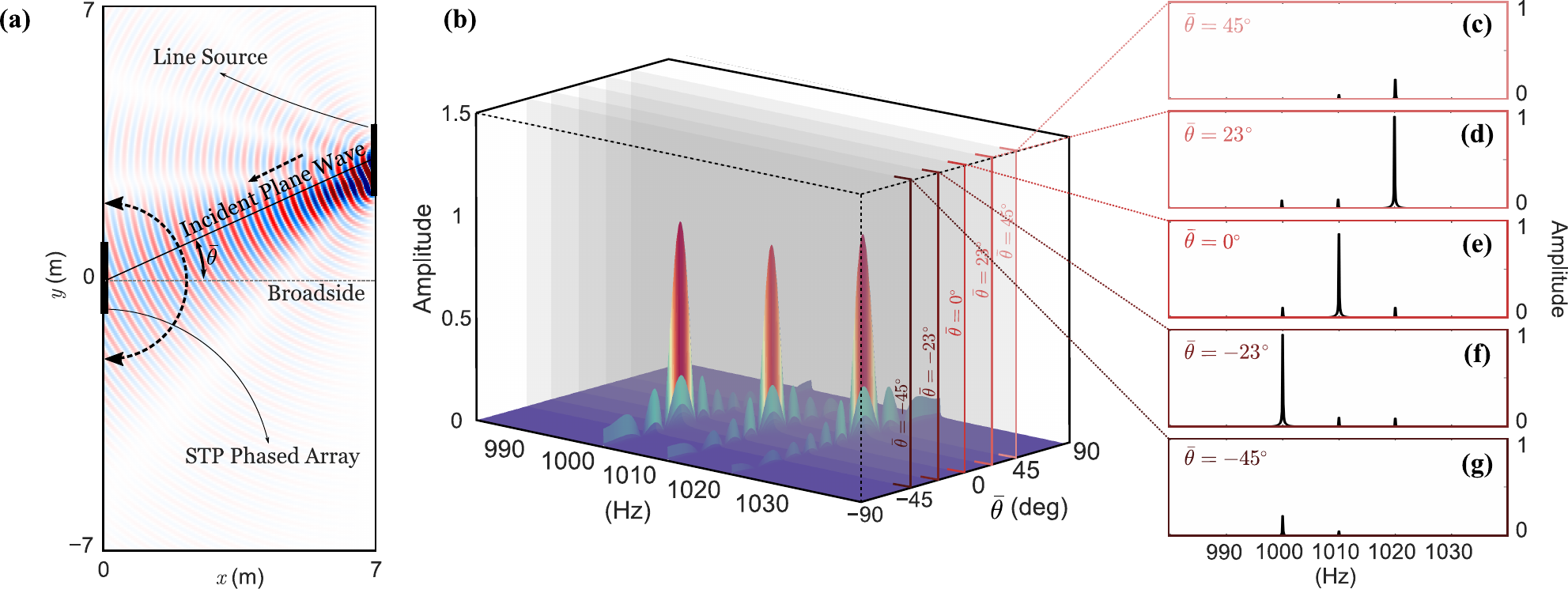}
\caption{(a) The STP phased array in $\mathbb{RX}$ mode with $\delta=1.5$. A wave is incident upon the array from an arbitrary angle $\bar{\theta}$ with a dominant frequency of $\bar{\omega}$. (b) FFT amplitude of the collected output signal for waves incident from $\bar{\theta}$ ranging from $-90^\circ$ to $90^\circ$ with $1^\circ$ increments. Parameters used are as follows: $\bar{\omega}/{2\pi} = 1010$ Hz and $\omega_m/2\pi=10$ Hz. (c-g) Slices of (b) at different incident directions: $\bar{\theta}=45^\circ$, $\bar{\theta}=\bar{\theta}_{s, +1} = 23^\circ$, $\bar{\theta}=\bar{\theta}_{s} = 0^\circ$, $\bar{\theta}=\bar{\theta}_{s, -1} = -23^\circ$, and $\bar{\theta}=-45^\circ$, respectively. We note that as the incident angles match either of the three listening directions of the array, a drastically higher voltage output can be detected which can be employed to determine the direction of arrival (DOA).}
\vspace{-0.5cm}
\label{fig:recpt}
\end{figure}

As described earlier, the STP phase shifters follow a traveling-wave-like variation.
Here we consider a temporal modulation frequency of $\frac{\omega_m}{2\pi} = 10$ Hz and a spatial modulation frequency of $\kappa_m = 2.3344\pi$ rad/m. As a result, the spatial super cell spans 10 successive transducers. This yields a traveling modulation velocity of $\nu_m = \frac{\omega_m}{\kappa_m} = 8.6$ m/s which is about $1/40$ the speed of sound in air, $c$.
In the $\mathbb{TX}$ mode, the phased array is provided with a voltage input at $\frac{\omega}{2\pi} = 1000$ Hz. Fig.~\ref{fig:matlab} shows the STP array's response in the time and frequency domains while operating in the $\mathbb{TX}$ mode where $\kappa_s$, and thus $\theta_s$, are set to zero for simpler visualization.
Fig.~\ref{fig:matlab}a illustrates the resultant acoustic pressure field at $t = 1$ s for $\delta=0$, where only the fundamental wave component propagates along the broadside; resembling a conventional phased array.
This is further confirmed by the Fast Fourier Transform (FFT) analysis in Fig.~\ref{fig:matlab}b which is carried out at the three distinct frequencies shown. Fig.~\ref{fig:matlab}d shows the same array when a space-time modulation is triggered by setting $\delta=1.5$. In addition to the fundamental wave component traveling along the broadside, down- and up-converted waves can now be observed propagating at $\omega^{(1-)}=990$ Hz and $\omega^{(1+)}=1010$ Hz along the $\theta_{s, -1}=-24^\circ$ and $\theta_{s, +1}=23^\circ$ directions, respectively. A visualization of the same is best illustrated in Fig.~\ref{fig:matlab}e, where an FFT separates the wave components by frequency content, highlighting their respective propagation directions. It is important to note that the propagation angles shown here are in excellent agreement with Eq.~\prettyref{eq:thetas}, which is derived using the far-field approximation. Figs.~\ref{fig:matlab}c and f show the amplitude of the wave components at three distinct pressure sensing locations denoted by the red, blue, and purple dots in the main figures.
These locations are selected at a radius of $15\lambda$ from the center of the phased array at $=-24^\circ$, $0^\circ$ and $23^\circ$ measured from the broadside, respectively.
Given the presence of three wave components with comparable amplitudes in the $\delta = 1.5$ case, we limit the rest of our analysis to this $\delta$ value.
In order to simulate the behavior of the STP phased array with $\delta=1.5$ in the $\mathbb{RX}$ mode, a plane wave line source is placed at a sufficiently far distance from the center of the array, and at $1^\circ$ angular increments spanning the range $\bar{\theta}=-90^\circ$ to $90^\circ$, while generating waves with a frequency of $\bar{\omega}=1010$ Hz as shown in Fig.~\ref{fig:recpt}a.
As explained earlier, the signal collected from the array shows a dominant amplitude at one of the, $\bar{\omega}$, $\bar{\omega}^{(1+)}$ or $\bar{\omega}^{(1-)}$ frequencies depending on the incident direction of the wave.
Exploiting this phenomenon enables a substantial multi-channel scanning capability of the free space as well as an ability to identify the direction of arrival (DOA) (Refer to the Supplementary Information for more on the application of multi-channel operation of an STP phased array).
Fig.~\ref{fig:recpt}b depicts the FFT of the resultant voltage output of the array when excited at different $\bar{\theta}$, which aligns very well with the coefficients of the fundamental, up- and down-converted signal terms $B_{0}$, $B_{+1}$, and $B_{-1}$ derived earlier.
As anticipated, the principal listening channels---i.e., where the peaks appear in Fig.~\ref{fig:recpt}---are in the fundamental $\bar{\theta} = 0^\circ$, up-converted $\bar{\theta}_{s, +1} = 23^\circ$, and down-converted $\bar{\theta}_{s, -1} = -23^\circ$ directions.
In other words, the simulations confirm that if a plane wave is incident from the direction of $\bar{\theta}=\bar{\theta}_{s, +1}$, the dominant frequency in the collected voltage signal becomes $\bar{\omega}^{(1+)}$.
Examples of this behavior are given in Fig.~\ref{fig:recpt}c-g. For instance, in Fig.~\ref{fig:recpt}d, the plane wave is incident with $\frac{\bar{\omega}}{2\pi} = 1010$ Hz while the dominant frequency in the collected signal is $\frac{\bar{\omega}^{(1+)}}{2\pi} = 1020$ Hz. 
As a result, we conclude that the DOA is $23^\circ$.
Similar arguments can be extended to Figs.~\ref{fig:recpt}e and f where the respective DOAs are found to correspond to $0^{\circ}$ and $-23^\circ$.
Once more, we emphasize that the principal directions of the array are \textit{in situ} tunable, which---in combination with the aforementioned scanning capacity---embody the potential of such arrays in the $\mathbb{RX}$ mode.

\section*{Numerical Validation and Methods}
\label{sec:comsol}
\begin{figure}[th!]
    \centering
    \includegraphics[width =\linewidth]{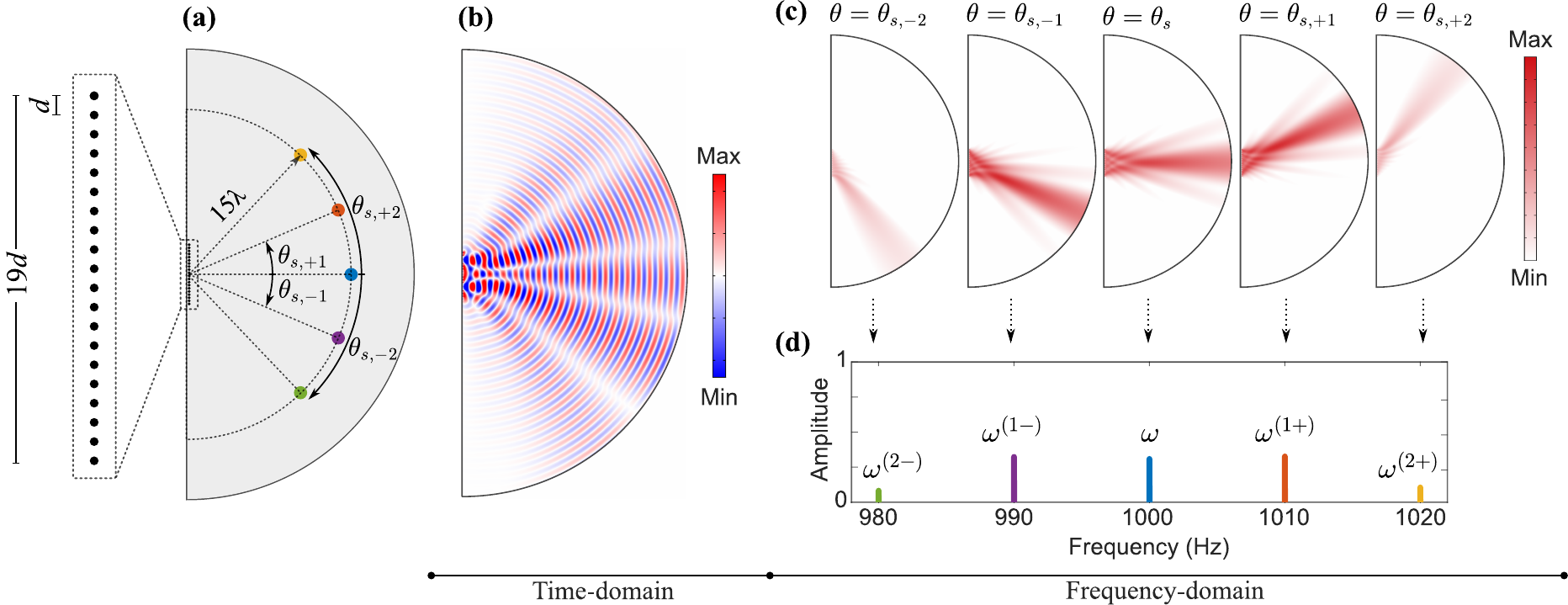}
    \caption{Time-domain finite element simulations of the STP phased array in $\mathbb{TX}$ mode for a modulation amplitude of $\delta = 1.5$. (a) A schematic diagram of the considered semicircle acoustic domain, dipole sources, and sensor locations. (b) Pressure field at $t=0.05$ s with $\omega/2\pi=1000$ Hz and $\omega_m/2\pi=10$ Hz. (c) From left to right: Directional breakdown and distribution of the FFT amplitudes at $980$ Hz, $990$ Hz,  $1000$ Hz, $1010$ Hz, and $1020$ Hz. (d) Normalized frequency spectrum of the pressure amplitude for the five sensor locations shown in (a). Parameters used are as follows: $\lambda=343$ mm, $d=\lambda/4$, and principal transmission directions are measured at $\theta_{s,-2}=-55^\circ$,  $\theta_{s,-1}=-24^\circ$, $\theta_{s}=0^\circ$, $\theta_{s,+1}=23^\circ$, and $\theta_{s,+2}=52^\circ$.}
    \label{fig:COMSOL_trans}
\end{figure}

In order to justify the first-order and far-field approximations exercised earlier, a number of highly computational finite element COMSOL simulations are herein carried out to assess these assumptions for both the $\mathbb{TX}$ and $\mathbb{RX}$ modes.
A two-dimensional acoustic domain comprising an air-filled semicircle with a $7$ m radius is considered. The speed of sound  $c$ is $343$ m/s and an air density of $\rho=1.2$ kg/m$^3$ is utilized.
Plane wave radiation boundary conditions are assigned to the surrounding walls to mitigate back-scattering and reflections of acoustic waves and reflections in order to reproduce the free-space behavior.
Similar to the previous experiment, 20 dipole acoustic sources are spaced at a quarter wavelength and used to create the linear STP phased array. The array is excited with the voltage signal given by Eq.~\prettyref{eq:vol_sig}. The same set of parameters listed earlier were used here.
A schematic of the model is depicted in Fig.~\ref{fig:COMSOL_trans}a, which includes two spatial super cells, each measuring $10d = 857.5$ mm in length.

Starting with the $\mathbb{TX}$ mode, the pressure field of the STP phased array is depicted in Fig.~\ref{fig:COMSOL_trans}b, exhibiting five different transmission channels for the generated components. In Fig.~\ref{fig:COMSOL_trans}c, the transient results are post-processed and a series of FFTs are computed which break down the frequency content of these wave components into various $\mathbb{TX}$ channels. The fundamental wave component is observed at $\frac{\omega}{2\pi}=1000$ Hz, the first up-converted at $\frac{\omega^{(1+)}}{2\pi}=1010$ Hz, the first down-converted at $\frac{\omega^{(1-)}}{2\pi}=990$ Hz, the second up-converted at $\frac{\omega^{(2+)}}{2\pi}=1020$ Hz, and the second down-converted at $\frac{\omega^{(2-)}}{2\pi}=980$ Hz,
which approximately propagate along the $\theta_{s}=0^\circ$, $\theta_{s,+1}=23^\circ$, $\theta_{s,-1}=-24^\circ$, $\theta_{s,+2}=52^\circ$, and $\theta_{s,-2}=-55^\circ$ directions, respectively.
The previous angles are in agreement with the principal transmission directions expressed in Eq.~\prettyref{eq:thetas}.
We also note that second order wave components now appear in Fig.~\ref{fig:COMSOL_trans} due to the fact that all orders of Bessel functions are inherently considered in the present numerical simulations.
The normalized FFT spectrum of the pressure amplitudes measured at the sensors marked on Fig.~\ref{fig:COMSOL_trans}a is shown in Fig.~\ref{fig:COMSOL_trans}d, which follows the same color key.
As evident in the figure, the smaller amplitudes of the second order waves justify neglecting them in the theoretical derivations.
This can be attributed to two facts: First, the values of second order Bessel functions are smaller than first order ones. Second, as we approach the end-fire axis, the array's directional effects become stronger and more pronounced, which further reduce the amplitudes of such second order components.

\begin{figure}[th!]
    \centering
    \includegraphics[width =\linewidth]{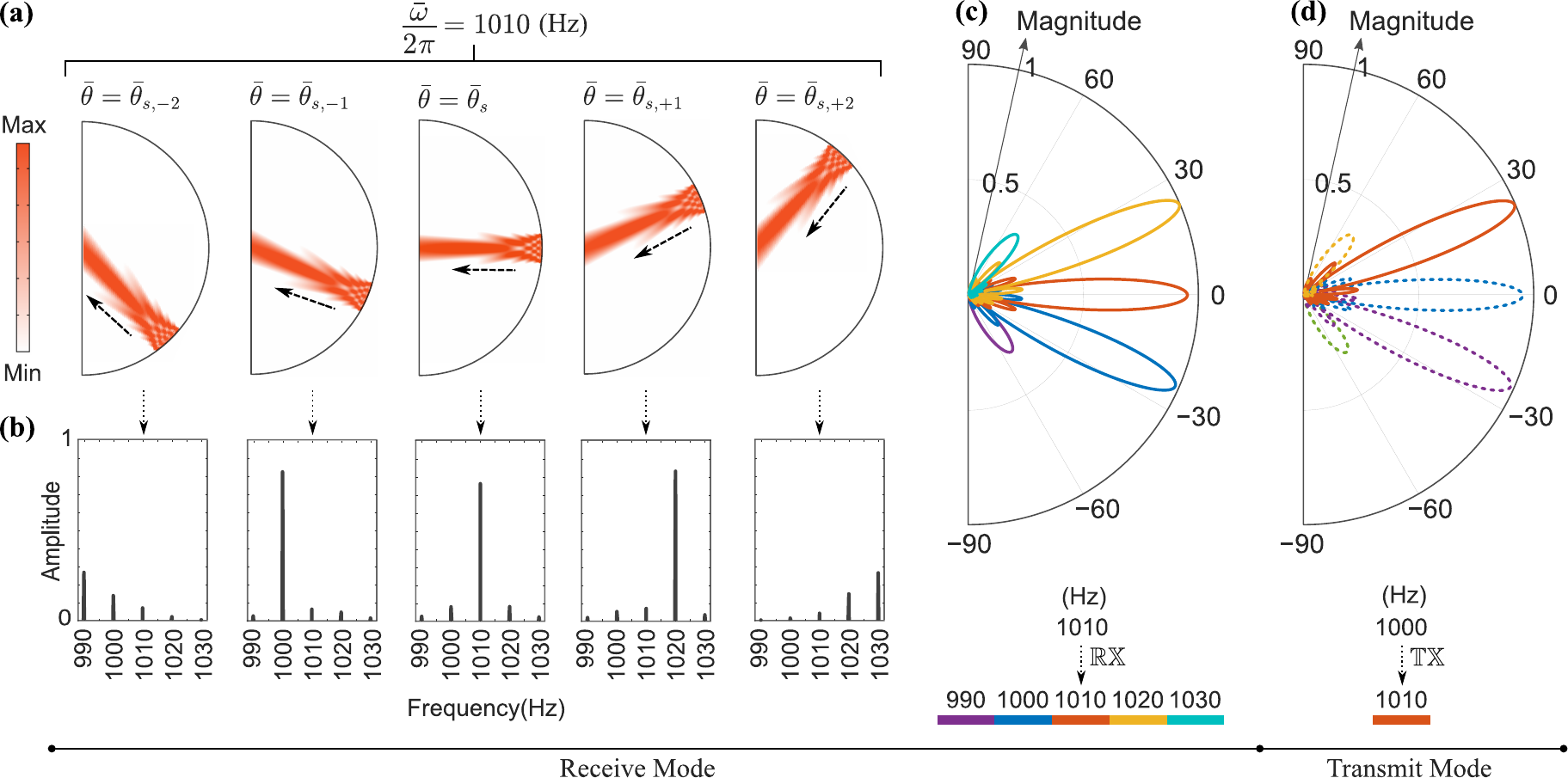}
    \caption{Time-domain finite element simulations of the STP phased array in $\mathbb{RX}$ mode for a modulation depth of $\delta = 1.5$. (a) Plane wave at $\frac{\bar{\omega}}{2\pi}=1010$ Hz incident upon the array from (left to right): $\bar{\theta}_{s,-2}=-55^\circ$,  $\bar{\theta}_{s,-1}=-23^\circ$, $\bar{\theta}_{s}=0^\circ$, $\bar{\theta}_{s,+1}=23^\circ$ and $\bar{\theta}_{s,+2}=55^\circ$. (b) Normalized FFT of the collected voltage signal amplitude for each case in (a). (c) Radiation pattern in $\mathbb{RX}$ mode for the different listening channels. (d) Radiation pattern in $\mathbb{TX}$ mode with waves propagating in three directions including the first up-converted component.}
    \label{fig:COMSOL_recep}
\end{figure}

The radiation pattern in the $\mathbb{RX}$ mode was also verified by sending acoustic waves at $\frac{\bar{\omega}}{2\pi}=1010$ Hz generated using a velocity line source from five specific incident angles:
$\bar{\theta}_{s,-2}=-55^\circ$,  $\bar{\theta}_{s,-1}=-23^\circ$, $\bar{\theta}_{s}=0^\circ$, $\bar{\theta}_{s,+1}=23^\circ$, and $\bar{\theta}_{s,+2}=55^\circ$ as depicted in Fig.~\ref{fig:COMSOL_recep}a.
We chose these angles since it is rather impractical to send incident plane waves towards the phased array from infinite distinct directions in a finite element model.
After passing through the space-time phase shifters, the received voltage signals at all the transducers accumulate a dynamic phase angle and are collected and summed up for each incident angle in Fig.~\ref{fig:COMSOL_recep}a. 
Following which, a series of FFTs are carried out on these signals and their normalized FFT spectra are presented in Fig.~\ref{fig:COMSOL_recep}b. 
It is evident that if the incoming plane wave is incident from $\bar{\theta}_{s,-2}$,  $\bar{\theta}_{s,-1}$, $\bar{\theta}_{s}$, $\bar{\theta}_{s,+1}$, and $\bar{\theta}_{s,+2}$, then the second down-converted ($\frac{\bar{\omega}^{(2-)}}{2\pi}=990$ Hz), first down-converted ($\frac{\bar{\omega}^{(1-)}}{2\pi}=1000$ Hz), fundamental ($\frac{\bar{\omega}}{2\pi}=1010$ Hz), first up-converted ($\frac{\bar{\omega}^{(1+)}}{2\pi}=1020$ Hz), and second up-converted ($\frac{\bar{\omega}^{(2+)}}{2\pi}=1030$ Hz) signal components will be most effectively detected. 
This agrees well with the principal listening directions derived in Eq.~(\ref{eq:barthetas}). 
As a result, in practical applications, the frequency ($\bar{f}$) and incident angle ($\bar{\theta}$) of the incoming beam can be figured out by inspecting the FFT amplitudes of the received voltages only.
Furthermore, radiation patterns in the $\mathbb{RX}$ mode for different listening channels are presented in Fig.~\ref{fig:COMSOL_recep}c, and show the sensitivity of the STP phased array to an incident plane wave of frequency $\frac{\bar{\omega}}{2\pi}=1010$ Hz as a function of incident angle. 
Finally, to correlate these patterns with the three nonreciprocity categories outlined earlier, the red solid line in Fig.~\ref{fig:COMSOL_recep}d shows the radiation pattern of the array in the $\mathbb{TX}$ mode for the first up-converted wave component.
Comparing Fig.~\ref{fig:COMSOL_recep}d with the listening channels of 1010 Hz $\rightarrow$ 1000 Hz, 1010 Hz $\rightarrow$ 1020 Hz, and 1010 Hz $\rightarrow$ 1010 Hz in Fig.~\ref{fig:COMSOL_recep}c, is indicative of reciprocity breakage in momentum, frequency, and both domains, respectively.

\section*{Summary}
\label{sec:conclusion}
A linear acoustic phased array was proposed, which provides independent control of transmission and reception patterns, opening up the possibility of nonreciprocal operation. Unlike its conventional counterpart, the space-time-periodic phased array is capable of generating additional side bands that carry higher and lower harmonics. The phased array comprises multiple phase-shifters and transducers which are paired up and stacked to form a subwavelength device. The phase-shifters are dynamically modulated to follow a space-time-periodic pattern with a modulation that travels relatively slower than the speed of sound.
The operational principle of the phased array was developed through theoretical derivation and a Jacobi-Anger series expansion. Additionally, we demonstrated the dual operation of the space-time-periodic phased array in both transmission and reception modes. Through multiple numerical simulations, various possible ways of breaking wave reciprocity have been illustrated and the control over the directivity of transmitted and received waves was demonstrated. The proposed phased array can be of great value to practical applications involving acoustic telecommunication, underwater navigation as well as sea bed research.

\section*{Acknowledgements}

M.N. acknowledges support of this work from the US National Science Foundation through awards no.~1847254 (CAREER) and 1904254. A.A. and M.N. acknowledge support from the University at Buffalo through the Buffalo Blue Sky Program.

\section*{Author contributions statement}

M.A.A. developed the concept. R.A., M.M. carried out the analysis. M.A.A. and M.N. analysed the results. A.A. and M.N. supervised the work. All authors wrote the manuscript. 

\section*{Additional information}

\textbf{Data availability} All data generated or analysed during this study are included in this article (and its Supplementary Information files); \textbf{Competing interests} The authors declare no competing interests.


\bibliography{references}

\begin{thebibliography}{10}
\urlstyle{rm}
\expandafter\ifx\csname url\endcsname\relax
  \def\url#1{\texttt{#1}}\fi
\expandafter\ifx\csname urlprefix\endcsname\relax\def\urlprefix{URL }\fi
\expandafter\ifx\csname doiprefix\endcsname\relax\def\doiprefix{DOI: }\fi
\providecommand{\bibinfo}[2]{#2}
\providecommand{\eprint}[2][]{\url{#2}}

\bibitem{beamsteering}
\bibinfo{author}{{Von Ramm}, O.~T.} \& \bibinfo{author}{{Smith}, S.~W.}
\newblock \bibinfo{journal}{\bibinfo{title}{Beam steering with linear arrays}}.
\newblock {\emph{\JournalTitle{IEEE Transactions on Biomedical Engineering}}}
  \textbf{\bibinfo{volume}{BME-30}}, \bibinfo{pages}{438--452}
  (\bibinfo{year}{1983}).

\bibitem{beamsteering1}
\bibinfo{author}{Wooh, S.-C.} \& \bibinfo{author}{Shi, Y.}
\newblock \bibinfo{journal}{\bibinfo{title}{Optimum beam steering of linear
  phased arrays}}.
\newblock {\emph{\JournalTitle{Wave Motion}}} \textbf{\bibinfo{volume}{29}},
  \bibinfo{pages}{245 -- 265} (\bibinfo{year}{1999}).

\bibitem{jones1988technical}
\bibinfo{author}{Jones, S.}
\newblock \bibinfo{journal}{\bibinfo{title}{Technical history of the beginnings
  of radar. ss swords. 342 pages, 23$\times$ 15 cm, peter perigrenus,
  1986.{\pounds} 39.00.}}
\newblock {\emph{\JournalTitle{The Journal of Navigation}}}
  \textbf{\bibinfo{volume}{41}}, \bibinfo{pages}{443--444}
  (\bibinfo{year}{1988}).

\bibitem{poulton_lidar}
\bibinfo{author}{Poulton, C.~V.} \emph{et~al.}
\newblock \bibinfo{journal}{\bibinfo{title}{Coherent solid-state lidar with
  silicon photonic optical phased arrays}}.
\newblock {\emph{\JournalTitle{Opt. Lett.}}} \textbf{\bibinfo{volume}{42}},
  \bibinfo{pages}{4091--4094} (\bibinfo{year}{{ts }, doi =
  {10.1364/OL.42.004091}}).

\bibitem{radar}
\bibinfo{author}{{Hassanien}, A.} \& \bibinfo{author}{{Vorobyov}, S.~A.}
\newblock \bibinfo{journal}{\bibinfo{title}{Phased-mimo radar: A tradeoff
  between phased-array and mimo radars}}.
\newblock {\emph{\JournalTitle{IEEE Transactions on Signal Processing}}}
  \textbf{\bibinfo{volume}{58}}, \bibinfo{pages}{3137--3151}
  (\bibinfo{year}{2010}).

\bibitem{jorgensen1993doppler}
\bibinfo{author}{Jorgensen, K.~V.}, \bibinfo{author}{Grose, B.~L.} \&
  \bibinfo{author}{Crandall, F.~A.}
\newblock \bibinfo{title}{Doppler sonar applied to precision underwater
  navigation}.
\newblock In \emph{\bibinfo{booktitle}{Proceedings of OCEANS'93}},
  \bibinfo{pages}{II469--II474} (\bibinfo{organization}{IEEE},
  \bibinfo{year}{1993}).

\bibitem{jensen2007medical}
\bibinfo{author}{Jensen, J.~A.}
\newblock \bibinfo{journal}{\bibinfo{title}{Medical ultrasound imaging}}.
\newblock {\emph{\JournalTitle{Progress in biophysics and molecular biology}}}
  \textbf{\bibinfo{volume}{93}}, \bibinfo{pages}{153--165}
  (\bibinfo{year}{2007}).

\bibitem{rynne1998phased}
\bibinfo{author}{Rynne, T.~M.}, \bibinfo{author}{Spadaro, J.~F.},
  \bibinfo{author}{Iovenitti, J.~L.}, \bibinfo{author}{Dering, J.~P.} \&
  \bibinfo{author}{Hill, D.~G.}
\newblock \bibinfo{title}{Phased array approach to retrieve gases, liquids, or
  solids from subaqueous geologic or man-made formations}
  (\bibinfo{year}{1998}).
\newblock \bibinfo{note}{US Patent 5,826,653}.

\bibitem{birtill1965application}
\bibinfo{author}{Birtill, J.} \& \bibinfo{author}{Whiteway, F.}
\newblock \bibinfo{journal}{\bibinfo{title}{The application of phased arrays to
  the analysis of seismic body waves}}.
\newblock {\emph{\JournalTitle{Philosophical Transactions of the Royal Society
  of London. Series A, Mathematical and Physical Sciences}}}
  \textbf{\bibinfo{volume}{258}}, \bibinfo{pages}{421--493}
  (\bibinfo{year}{1965}).

\bibitem{mcnab1987ultrasonic}
\bibinfo{author}{McNab, A.} \& \bibinfo{author}{Campbell, M.}
\newblock \bibinfo{journal}{\bibinfo{title}{Ultrasonic phased arrays for
  nondestructive testing}}.
\newblock {\emph{\JournalTitle{NDT international}}}
  \textbf{\bibinfo{volume}{20}}, \bibinfo{pages}{333--337}
  (\bibinfo{year}{1987}).

\bibitem{zhou_2016}
\bibinfo{author}{Zhou, Y.-Q.} \& \bibinfo{author}{Zhan, L.-H.}
\newblock \bibinfo{title}{Application of ultrasonic phased array technology to
  the detection of defect in composite stiffened-structures}.
\newblock In \emph{\bibinfo{booktitle}{Advanced Material Science and
  Engineering}}, \bibinfo{pages}{315--322} (\bibinfo{publisher}{World
  Scientific}, \bibinfo{year}{2016}).

\bibitem{marzo_virtualvortex}
\bibinfo{author}{Marzo, A.}, \bibinfo{author}{Caleap, M.} \&
  \bibinfo{author}{Drinkwater, B.~W.}
\newblock \bibinfo{journal}{\bibinfo{title}{Acoustic virtual vortices with
  tunable orbital angular momentum for trapping of mie particles}}.
\newblock {\emph{\JournalTitle{Phys. Rev. Lett.}}}
  \textbf{\bibinfo{volume}{120}}, \bibinfo{pages}{044301},
  \doiprefix\url{10.1103/PhysRevLett.120.044301} (\bibinfo{year}{2018}).

\bibitem{hoshi_2014}
\bibinfo{author}{Hoshi, T.}, \bibinfo{author}{Ochiai, Y.} \&
  \bibinfo{author}{Rekimoto, J.}
\newblock \bibinfo{journal}{\bibinfo{title}{Three-dimensional noncontact
  manipulation by opposite ultrasonic phased arrays}}.
\newblock {\emph{\JournalTitle{Japanese Journal of Applied Physics}}}
  \textbf{\bibinfo{volume}{53}}, \bibinfo{pages}{07KE07},
  \doiprefix\url{10.7567/JJAP.53.07KE07} (\bibinfo{year}{2014}).

\bibitem{marzo_tweezers}
\bibinfo{author}{Marzo, A.} \& \bibinfo{author}{Drinkwater, B.~W.}
\newblock \bibinfo{journal}{\bibinfo{title}{Holographic acoustic tweezers}}.
\newblock {\emph{\JournalTitle{Proceedings of the National Academy of
  Sciences}}} \textbf{\bibinfo{volume}{116}}, \bibinfo{pages}{84--89},
  \doiprefix\url{10.1073/pnas.1813047115} (\bibinfo{year}{2019}).
\newblock \eprint{https://www.pnas.org/content/116/1/84.full.pdf}.

\bibitem{starlink1}
\bibinfo{author}{{Sayin}, A.}, \bibinfo{author}{{Cherniakov}, M.} \&
  \bibinfo{author}{{Antoniou}, M.}
\newblock \bibinfo{title}{Passive radar using starlink transmissions: A
  theoretical study}.
\newblock In \emph{\bibinfo{booktitle}{2019 20th International Radar Symposium
  (IRS)}}, \bibinfo{pages}{1--7} (\bibinfo{year}{2019}).

\bibitem{starlink2}
\bibinfo{author}{Amato, F.} \emph{et~al.}
\newblock \bibinfo{title}{{Photonic integrated circuits for ultra-fast steering
  in phased-array antennas}}.
\newblock In \bibinfo{editor}{Sodnik, Z.}, \bibinfo{editor}{Karafolas, N.} \&
  \bibinfo{editor}{Cugny, B.} (eds.) \emph{\bibinfo{booktitle}{International
  Conference on Space Optics — ICSO 2018}}, vol. \bibinfo{volume}{11180},
  \bibinfo{pages}{2601 -- 2609}, \doiprefix\url{10.1117/12.2536179}.
  \bibinfo{organization}{International Society for Optics and Photonics}
  (\bibinfo{publisher}{SPIE}, \bibinfo{year}{2019}).

\bibitem{bashri}
\bibinfo{author}{{Bashri}, M. S.~R.}, \bibinfo{author}{{Arslan}, T.} \&
  \bibinfo{author}{{Zhou}, W.}
\newblock \bibinfo{title}{A dual-band linear phased array antenna for wifi and
  lte mobile applications}.
\newblock In \emph{\bibinfo{booktitle}{2015 Loughborough Antennas Propagation
  Conference (LAPC)}}, \bibinfo{pages}{1--5} (\bibinfo{year}{2015}).

\bibitem{qian}
\bibinfo{author}{{Qian}, K.} \emph{et~al.}
\newblock \bibinfo{journal}{\bibinfo{title}{Enabling phased array signal
  processing for mobile wifi devices}}.
\newblock {\emph{\JournalTitle{IEEE Transactions on Mobile Computing}}}
  \textbf{\bibinfo{volume}{17}}, \bibinfo{pages}{1820--1833}
  (\bibinfo{year}{2018}).

\bibitem{weather}
\bibinfo{author}{Zrnic, D.~S.} \emph{et~al.}
\newblock \bibinfo{journal}{\bibinfo{title}{Agile-beam phased array radar for
  weather observations}}.
\newblock {\emph{\JournalTitle{Bulletin of the American Meteorological
  Society}}} \textbf{\bibinfo{volume}{88}}, \bibinfo{pages}{1753--1766},
  \doiprefix\url{10.1175/BAMS-88-11-1753} (\bibinfo{year}{2007}).
\newblock \eprint{https://doi.org/10.1175/BAMS-88-11-1753}.

\bibitem{arnon:s}
\bibinfo{author}{Arnon, S.}, \bibinfo{author}{Rotman, S.} \&
  \bibinfo{author}{Kopeika, N.~S.}
\newblock \bibinfo{journal}{\bibinfo{title}{Beam width and transmitter power
  adaptive to tracking system performance for free-space optical
  communication}}.
\newblock {\emph{\JournalTitle{Appl. Opt.}}} \textbf{\bibinfo{volume}{36}},
  \bibinfo{pages}{6095--6101}, \doiprefix\url{10.1364/AO.36.006095}
  (\bibinfo{year}{1997}).

\bibitem{assouar2018acoustic}
\bibinfo{author}{Assouar, B.} \emph{et~al.}
\newblock \bibinfo{journal}{\bibinfo{title}{Acoustic metasurfaces}}.
\newblock {\emph{\JournalTitle{Nature Reviews Materials}}}
  \textbf{\bibinfo{volume}{3}}, \bibinfo{pages}{460--472}
  (\bibinfo{year}{2018}).

\bibitem{hansen2009phased}
\bibinfo{author}{Hansen, R.~C.}
\newblock \emph{\bibinfo{title}{Phased array antennas}}, vol.
  \bibinfo{volume}{213} (\bibinfo{publisher}{John Wiley \& Sons},
  \bibinfo{year}{2009}).

\bibitem{achenbach2003reciprocity}
\bibinfo{author}{Achenbach, J.} \& \bibinfo{author}{Achenbach, J.}
\newblock \emph{\bibinfo{title}{Reciprocity in elastodynamics}}
  (\bibinfo{publisher}{Cambridge University Press}, \bibinfo{year}{2003}).

\bibitem{lorentz1896theorem}
\bibinfo{author}{Lorentz, H.~A.}
\newblock \bibinfo{journal}{\bibinfo{title}{The theorem of poynting concerning
  the energy in the electromagnetic field and two general propositions
  concerning the propagation of light}}.
\newblock {\emph{\JournalTitle{Amsterdammer Akademie der Wetenschappen}}}
  \textbf{\bibinfo{volume}{4}}, \bibinfo{pages}{176} (\bibinfo{year}{1896}).

\bibitem{attarzadeh-2018}
\bibinfo{author}{Attarzadeh, M.}, \bibinfo{author}{AlBa’ba’a, H.} \&
  \bibinfo{author}{Nouh, M.}
\newblock \bibinfo{journal}{\bibinfo{title}{On the wave dispersion and
  non-reciprocal power flow in space-time traveling acoustic metamaterials}}.
\newblock {\emph{\JournalTitle{Applied Acoustics}}}
  \textbf{\bibinfo{volume}{133}}, \bibinfo{pages}{210 -- 214},
  \doiprefix\url{https://doi.org/10.1016/j.apacoust.2017.12.028}
  (\bibinfo{year}{2018}).

\bibitem{nassar2017non}
\bibinfo{author}{Nassar, H.}, \bibinfo{author}{Chen, H.},
  \bibinfo{author}{Norris, A.} \& \bibinfo{author}{Huang, G.}
\newblock \bibinfo{journal}{\bibinfo{title}{Non-reciprocal flexural wave
  propagation in a modulated metabeam}}.
\newblock {\emph{\JournalTitle{Extreme Mechanics Letters}}}
  \textbf{\bibinfo{volume}{15}}, \bibinfo{pages}{97--102}
  (\bibinfo{year}{2017}).

\bibitem{chen2020active}
\bibinfo{author}{Chen, Y.}, \bibinfo{author}{Li, X.}, \bibinfo{author}{Hu, G.},
  \bibinfo{author}{Haberman, M.~R.} \& \bibinfo{author}{Huang, G.}
\newblock \bibinfo{journal}{\bibinfo{title}{An active mechanical willis
  meta-layer with asymmetric polarizabilities}}.
\newblock {\emph{\JournalTitle{Nature communications}}}
  \textbf{\bibinfo{volume}{11}}, \bibinfo{pages}{1--8} (\bibinfo{year}{2020}).

\bibitem{li2011tunable}
\bibinfo{author}{Li, X.-F.} \emph{et~al.}
\newblock \bibinfo{journal}{\bibinfo{title}{Tunable unidirectional sound
  propagation through a sonic-crystal-based acoustic diode}}.
\newblock {\emph{\JournalTitle{Physical review letters}}}
  \textbf{\bibinfo{volume}{106}}, \bibinfo{pages}{084301}
  (\bibinfo{year}{2011}).

\bibitem{boechler2011bifurcation}
\bibinfo{author}{Boechler, N.}, \bibinfo{author}{Theocharis, G.} \&
  \bibinfo{author}{Daraio, C.}
\newblock \bibinfo{journal}{\bibinfo{title}{Bifurcation-based acoustic
  switching and rectification}}.
\newblock {\emph{\JournalTitle{Nature materials}}}
  \textbf{\bibinfo{volume}{10}}, \bibinfo{pages}{665--668}
  (\bibinfo{year}{2011}).

\bibitem{yang2015topological}
\bibinfo{author}{Yang, Z.} \emph{et~al.}
\newblock \bibinfo{journal}{\bibinfo{title}{Topological acoustics}}.
\newblock {\emph{\JournalTitle{Physical review letters}}}
  \textbf{\bibinfo{volume}{114}}, \bibinfo{pages}{114301}
  (\bibinfo{year}{2015}).

\bibitem{nassar2020nonreciprocity}
\bibinfo{author}{Nassar, H.} \emph{et~al.}
\newblock \bibinfo{journal}{\bibinfo{title}{Nonreciprocity in acoustic and
  elastic materials}}.
\newblock {\emph{\JournalTitle{Nature Reviews Materials}}}
  \bibinfo{pages}{1--19} (\bibinfo{year}{2020}).

\bibitem{adam2002ferrite}
\bibinfo{author}{Adam, J.~D.}, \bibinfo{author}{Davis, L.~E.},
  \bibinfo{author}{Dionne, G.~F.}, \bibinfo{author}{Schloemann, E.~F.} \&
  \bibinfo{author}{Stitzer, S.~N.}
\newblock \bibinfo{journal}{\bibinfo{title}{Ferrite devices and materials}}.
\newblock {\emph{\JournalTitle{IEEE Transactions on Microwave Theory and
  Techniques}}} \textbf{\bibinfo{volume}{50}}, \bibinfo{pages}{721--737}
  (\bibinfo{year}{2002}).

\bibitem{sounas2017non}
\bibinfo{author}{Sounas, D.~L.} \& \bibinfo{author}{Al{\`u}, A.}
\newblock \bibinfo{journal}{\bibinfo{title}{Non-reciprocal photonics based on
  time modulation}}.
\newblock {\emph{\JournalTitle{Nature Photonics}}}
  \textbf{\bibinfo{volume}{11}}, \bibinfo{pages}{774--783}
  (\bibinfo{year}{2017}).

\bibitem{shupe1980thermally}
\bibinfo{author}{Shupe, D.~M.}
\newblock \bibinfo{journal}{\bibinfo{title}{Thermally induced nonreciprocity in
  the fiber-optic interferometer}}.
\newblock {\emph{\JournalTitle{Applied optics}}} \textbf{\bibinfo{volume}{19}},
  \bibinfo{pages}{654--655} (\bibinfo{year}{1980}).

\bibitem{hadad2013one}
\bibinfo{author}{Hadad, Y.} \& \bibinfo{author}{Steinberg, B.~Z.}
\newblock \bibinfo{journal}{\bibinfo{title}{One way optical waveguides for
  matched non-reciprocal nanoantennas with dynamic beam scanning
  functionality}}.
\newblock {\emph{\JournalTitle{Optics express}}} \textbf{\bibinfo{volume}{21}},
  \bibinfo{pages}{A77--A83} (\bibinfo{year}{2013}).

\bibitem{zang2019nonreciprocal}
\bibinfo{author}{Zang, J.}, \bibinfo{author}{Alvarez-Melcon, A.} \&
  \bibinfo{author}{Gomez-Diaz, J.}
\newblock \bibinfo{journal}{\bibinfo{title}{Nonreciprocal phased-array
  antennas}}.
\newblock {\emph{\JournalTitle{Physical Review Applied}}}
  \textbf{\bibinfo{volume}{12}}, \bibinfo{pages}{054008}
  (\bibinfo{year}{2019}).

\bibitem{guo2019nonreciprocal}
\bibinfo{author}{Guo, X.}, \bibinfo{author}{Ding, Y.}, \bibinfo{author}{Duan,
  Y.} \& \bibinfo{author}{Ni, X.}
\newblock \bibinfo{journal}{\bibinfo{title}{Nonreciprocal metasurface with
  space--time phase modulation}}.
\newblock {\emph{\JournalTitle{Light: Science \& Applications}}}
  \textbf{\bibinfo{volume}{8}}, \bibinfo{pages}{1--9} (\bibinfo{year}{2019}).

\bibitem{cardin2020surface}
\bibinfo{author}{Cardin, A.~E.} \emph{et~al.}
\newblock \bibinfo{journal}{\bibinfo{title}{Surface-wave-assisted
  nonreciprocity in spatio-temporally modulated metasurfaces}}.
\newblock {\emph{\JournalTitle{Nature communications}}}
  \textbf{\bibinfo{volume}{11}}, \bibinfo{pages}{1--9} (\bibinfo{year}{2020}).

\bibitem{zhang2018space}
\bibinfo{author}{Zhang, L.} \emph{et~al.}
\newblock \bibinfo{journal}{\bibinfo{title}{Space-time-coding digital
  metasurfaces}}.
\newblock {\emph{\JournalTitle{Nature communications}}}
  \textbf{\bibinfo{volume}{9}}, \bibinfo{pages}{1--11} (\bibinfo{year}{2018}).

\bibitem{zhang2019breaking}
\bibinfo{author}{Zhang, L.} \emph{et~al.}
\newblock \bibinfo{journal}{\bibinfo{title}{Breaking reciprocity with
  space-time-coding digital metasurfaces}}.
\newblock {\emph{\JournalTitle{Advanced Materials}}}
  \textbf{\bibinfo{volume}{31}}, \bibinfo{pages}{1904069}
  (\bibinfo{year}{2019}).

\bibitem{wang2020theory}
\bibinfo{author}{Wang, X.}, \bibinfo{author}{D{\'\i}az-Rubio, A.},
  \bibinfo{author}{Li, H.}, \bibinfo{author}{Tretyakov, S.~A.} \&
  \bibinfo{author}{Al{\`u}, A.}
\newblock \bibinfo{journal}{\bibinfo{title}{Theory and design of
  multifunctional space-time metasurfaces}}.
\newblock {\emph{\JournalTitle{Physical Review Applied}}}
  \textbf{\bibinfo{volume}{13}}, \bibinfo{pages}{044040}
  (\bibinfo{year}{2020}).

\bibitem{maznev2013reciprocity}
\bibinfo{author}{Maznev, A.}, \bibinfo{author}{Every, A.} \&
  \bibinfo{author}{Wright, O.}
\newblock \bibinfo{journal}{\bibinfo{title}{Reciprocity in reflection and
  transmission: What is a ‘phonon diode’?}}
\newblock {\emph{\JournalTitle{Wave Motion}}} \textbf{\bibinfo{volume}{50}},
  \bibinfo{pages}{776--784} (\bibinfo{year}{2013}).

\bibitem{fleury2015nonreciprocal}
\bibinfo{author}{Fleury, R.}, \bibinfo{author}{Sounas, D.},
  \bibinfo{author}{Haberman, M.~R.} \& \bibinfo{author}{Alu, A.}
\newblock \bibinfo{journal}{\bibinfo{title}{Nonreciprocal acoustics}}.
\newblock {\emph{\JournalTitle{Acoustics Today}}}
  \textbf{\bibinfo{volume}{11}}, \bibinfo{pages}{14--21}
  (\bibinfo{year}{2015}).

\bibitem{devaux2019acoustic}
\bibinfo{author}{Devaux, T.}, \bibinfo{author}{Cebrecos, A.},
  \bibinfo{author}{Richoux, O.}, \bibinfo{author}{Pagneux, V.} \&
  \bibinfo{author}{Tournat, V.}
\newblock \bibinfo{journal}{\bibinfo{title}{Acoustic radiation pressure for
  nonreciprocal transmission and switch effects}}.
\newblock {\emph{\JournalTitle{Nature communications}}}
  \textbf{\bibinfo{volume}{10}}, \bibinfo{pages}{1--8} (\bibinfo{year}{2019}).

\bibitem{moore2018nonreciprocity}
\bibinfo{author}{Moore, K.~J.} \emph{et~al.}
\newblock \bibinfo{journal}{\bibinfo{title}{Nonreciprocity in the dynamics of
  coupled oscillators with nonlinearity, asymmetry, and scale hierarchy}}.
\newblock {\emph{\JournalTitle{Physical Review E}}}
  \textbf{\bibinfo{volume}{97}}, \bibinfo{pages}{012219}
  (\bibinfo{year}{2018}).

\bibitem{popa2014non}
\bibinfo{author}{Popa, B.-I.} \& \bibinfo{author}{Cummer, S.~A.}
\newblock \bibinfo{journal}{\bibinfo{title}{Non-reciprocal and highly nonlinear
  active acoustic metamaterials}}.
\newblock {\emph{\JournalTitle{Nature communications}}}
  \textbf{\bibinfo{volume}{5}}, \bibinfo{pages}{1--5} (\bibinfo{year}{2014}).

\bibitem{liang2010acoustic}
\bibinfo{author}{Liang, B.}, \bibinfo{author}{Guo, X.}, \bibinfo{author}{Tu,
  J.}, \bibinfo{author}{Zhang, D.} \& \bibinfo{author}{Cheng, J.}
\newblock \bibinfo{journal}{\bibinfo{title}{An acoustic rectifier}}.
\newblock {\emph{\JournalTitle{Nature materials}}}
  \textbf{\bibinfo{volume}{9}}, \bibinfo{pages}{989--992}
  (\bibinfo{year}{2010}).

\bibitem{liang2009acoustic}
\bibinfo{author}{Liang, B.}, \bibinfo{author}{Yuan, B.} \&
  \bibinfo{author}{Cheng, J.-c.}
\newblock \bibinfo{journal}{\bibinfo{title}{Acoustic diode: Rectification of
  acoustic energy flux in one-dimensional systems}}.
\newblock {\emph{\JournalTitle{Physical review letters}}}
  \textbf{\bibinfo{volume}{103}}, \bibinfo{pages}{104301}
  (\bibinfo{year}{2009}).

\bibitem{attarzadeh2018}
\bibinfo{author}{Attarzadeh, M.~A.} \& \bibinfo{author}{Nouh, M.}
\newblock \bibinfo{journal}{\bibinfo{title}{Elastic wave propagation in moving
  phononic crystals and correlations with stationary spatiotemporally modulated
  systems}}.
\newblock {\emph{\JournalTitle{AIP Advances}}} \textbf{\bibinfo{volume}{8}},
  \bibinfo{pages}{105302}, \doiprefix\url{10.1063/1.5042252}
  (\bibinfo{year}{2018}).

\bibitem{fleury2014sound}
\bibinfo{author}{Fleury, R.}, \bibinfo{author}{Sounas, D.~L.},
  \bibinfo{author}{Sieck, C.~F.}, \bibinfo{author}{Haberman, M.~R.} \&
  \bibinfo{author}{Al{\`u}, A.}
\newblock \bibinfo{journal}{\bibinfo{title}{Sound isolation and giant linear
  nonreciprocity in a compact acoustic circulator}}.
\newblock {\emph{\JournalTitle{Science}}} \textbf{\bibinfo{volume}{343}},
  \bibinfo{pages}{516--519} (\bibinfo{year}{2014}).

\bibitem{fleury2015subwavelength}
\bibinfo{author}{Fleury, R.}, \bibinfo{author}{Sounas, D.~L.} \&
  \bibinfo{author}{Al{\`u}, A.}
\newblock \bibinfo{journal}{\bibinfo{title}{Subwavelength ultrasonic circulator
  based on spatiotemporal modulation}}.
\newblock {\emph{\JournalTitle{Physical Review B}}}
  \textbf{\bibinfo{volume}{91}}, \bibinfo{pages}{174306}
  (\bibinfo{year}{2015}).

\bibitem{attarzadeh2018non}
\bibinfo{author}{Attarzadeh, M.} \& \bibinfo{author}{Nouh, M.}
\newblock \bibinfo{journal}{\bibinfo{title}{Non-reciprocal elastic wave
  propagation in 2d phononic membranes with spatiotemporally varying material
  properties}}.
\newblock {\emph{\JournalTitle{Journal of Sound and Vibration}}}
  \textbf{\bibinfo{volume}{422}}, \bibinfo{pages}{264--277}
  (\bibinfo{year}{2018}).

\bibitem{palermo2020surface}
\bibinfo{author}{Palermo, A.}, \bibinfo{author}{Celli, P.},
  \bibinfo{author}{Yousefzadeh, B.}, \bibinfo{author}{Daraio, C.} \&
  \bibinfo{author}{Marzani, A.}
\newblock \bibinfo{journal}{\bibinfo{title}{Surface wave non-reciprocity via
  time-modulated metamaterials}}.
\newblock {\emph{\JournalTitle{arXiv preprint arXiv:2007.10531}}}
  (\bibinfo{year}{2020}).

\bibitem{wang2013optical}
\bibinfo{author}{Wang, D.-W.} \emph{et~al.}
\newblock \bibinfo{journal}{\bibinfo{title}{Optical diode made from a moving
  photonic crystal}}.
\newblock {\emph{\JournalTitle{Physical review letters}}}
  \textbf{\bibinfo{volume}{110}}, \bibinfo{pages}{093901}
  (\bibinfo{year}{2013}).

\bibitem{attarzadeh2020beam}
\bibinfo{author}{Attarzadeh, M.}, \bibinfo{author}{Callanan, J.} \&
  \bibinfo{author}{Nouh, M.}
\newblock \bibinfo{journal}{\bibinfo{title}{Experimental observation of
  nonreciprocal waves in a resonant metamaterial beam}}.
\newblock {\emph{\JournalTitle{Phys. Rev. Applied}}}
  \textbf{\bibinfo{volume}{13}}, \bibinfo{pages}{021001},
  \doiprefix\url{10.1103/PhysRevApplied.13.021001} (\bibinfo{year}{2020}).

\bibitem{balanis2016antenna}
\bibinfo{author}{Balanis, C.~A.}
\newblock \emph{\bibinfo{title}{Antenna theory: analysis and design}}
  (\bibinfo{publisher}{John wiley \& sons}, \bibinfo{year}{2016}).

\bibitem{kinsler2009}
\bibinfo{author}{Kinsler, L.}, \bibinfo{author}{Frey, A.},
  \bibinfo{author}{Coppens, A.} \& \bibinfo{author}{Sanders, J.}
\newblock \emph{\bibinfo{title}{Fundamentals of Acoustics, 4TH ED}}
  (\bibinfo{publisher}{Wiley India Pvt. Limited}, \bibinfo{year}{2009}).

\end{thebibliography}
\end{document}